\documentclass{emulateapj}
\usepackage{apjfonts}
\journalinfo{The Astrophysical Journal, 2009, in press}

\shorttitle{A PULSATIONAL MECHANISM FOR Be-STAR DISKS}
\shortauthors{STEVEN R. CRANMER}

\begin{document}

\title{A Pulsational Mechanism for Producing Keplerian Disks
around Be Stars}

\author{Steven R. Cranmer}
\affil{Harvard-Smithsonian Center for Astrophysics,
60 Garden Street, Cambridge, MA 02138}

\begin{abstract}
Classical Be stars are an enigmatic subclass of rapidly rotating
hot stars characterized by dense equatorial disks of gas that have
been inferred to orbit with Keplerian velocities.
Although these disks seem to be ejected from the star and not
accreted, there is substantial observational evidence to show that
the stars rotate more slowly than required for centrifugally
driven mass loss.
This paper develops an idea (proposed originally by Hiroyasu
Ando and colleagues) that nonradial stellar pulsations inject
enough angular momentum into the upper atmosphere to spin up a
Keplerian disk.
The pulsations themselves are evanescent in the stellar photosphere,
but they may be unstable to the generation of resonant oscillations
at the acoustic cutoff frequency.
A detailed theory of the conversion from pulsations to resonant
waves does not yet exist for realistic hot-star atmospheres, so
the current models depend on a parameterized approximation for
the efficiency of wave excitation.
Once resonant waves have been formed, however, they grow in
amplitude with increasing height, steepen into shocks, and exert
radial and azimuthal Reynolds stresses on the mean fluid.
Using reasonable assumptions for the stellar parameters, these
processes were found to naturally create the inner boundary
conditions required for dense Keplerian disks, even when the
underlying B-star photosphere is rotating as slowly as 60\% of
its critical rotation speed.
Because there is evidence for long-term changes in Be-star
pulsational properties, this model may also account for the
long-term variability of Be stars, including transitions between
normal, Be, and shell phases.
\end{abstract}

\keywords{circumstellar matter ---
stars: early-type ---
stars: oscillations (including pulsations) ---
stars: emission-line, Be ---
stars: rotation ---
waves}

\section{Introduction}

Be stars are non-supergiant B-type stars that exhibit emission
in their hydrogen Balmer lines.
It has been known for many decades (e.g., Struve 1931) that the
``classical'' Be stars tend to rotate more rapidly than normal
non-emission B stars.
A wide range of observations of Be stars is consistent with
the coexistence of a dense circumstellar disk in the equatorial
plane and a variable stellar wind at higher latitudes
(Doazan 1982; Slettebak 1988; Prinja 1989; Porter \& Rivinius 2003).
There is a great deal of evidence that the so-called
{\em decretion disk} is ejected from the star and is not accreted
from an external source of gas.
The physical mechanisms that are responsible for producing the
disk, though, are still not known.
This paper investigates the idea that nonradial pulsations (NRP)
may deposit sufficient angular momentum above the photospheres of
rapidly rotating Be stars to explain the origins of dense
equatorial disks.

Theoretical explanations for the Be phenomenon depend crucially
on how rapidly the stars and disks are rotating.
Although there is increasing evidence that the extended disks are
in Keplerian orbit (e.g., Hanuschik 1996; Hummel \& Vrancken 2000;
Okazaki 2007),
there is also evidence to show that the underlying photospheres
rotate too slowly to propel any atmospheric gas into orbit.
Typical observational values of the ratio of equatorial rotation
speed to the critical rotation speed $V_{\rm crit}$ (at which
gravity is balanced by the outward centrifugal force) range
between 0.5 and 0.8 (Slettebak 1982; Porter 1996; Yudin 2001).
There has been some recent disagreement between this traditional
picture and new ideas---inspired by the tendency for {\em gravity
darkening}\footnote{%
An oblate, rigidly rotating star is expected to undergo an internal
redistribution of its net radiative flux in proportion to the
magnitude of the centrifugally-modified gravity, which produces
hotter and brighter poles and a cooler, dimmer equator (see, e.g.,
von Zeipel 1924; Slettebak 1949; Collins 1963; Maeder 1999).}
to mask rapid rotation at the equator---that most or all Be stars
may be rotating very nearly at $V_{\rm crit}$ (Townsend et al.\  2004).
Cranmer (2005) found that this may be true for the late-type Be
stars (i.e., spectral types B3 and later), but early-type Be
stars do seem to be consistent with a range of intrinsic rotation
speeds between $\sim$40\% and 100\% critical.
For stars rotating sufficiently below $V_{\rm crit}$, any physical
model for the origin of Be-star disks requires a significant
increase in angular momentum above the photosphere.

This paper develops a set of ideas concerning how stellar NRP
may give rise to outwardly propagating circumstellar waves, which
in turn can deposit angular momentum in Be-star disks.
The history of this theoretical perspective is discussed in more
detail in {\S}~2.
However, a number of other disk formation mechanisms have been
proposed, and it is important to keep in mind that in some cases
these could be acting in cooperation or competition with the
pulsational processes advocated here.
Some of the other suggested mechanisms have been:
(1) stellar ``wind compression'' that may channel supersonically
outflowing gas from mid-latitudes to the equatorial plane by
conservation of angular momentum (Bjorkman \& Cassinelli 1993;
Bjorkman 2000),
(2) episodic ejections from some point on the star, with some
material being propelled forward into orbit and some propelled
backward to fall onto the star (Kroll \& Hanuschik 1997;
Owocki \& Cranmer 2002), and
(3) magnetic forces that can channel gas toward the equatorial
plane and provide sufficient torque to spin up a disk (e.g.,
Poe \& Friend 1986; Cassinelli et al.\  2002;
Brown et al.\  2004, 2008; Ud-Doula et al.\  2008).

The three mechanisms listed above have one general feature in
common: they depend on the existence of large-scale supersonic
flows in the circumstellar regions.
Thus, these processes appear likely to give rise to strong
variability on short (dynamical) time scales.
Although many Be stars do exhibit variability on time scales of
hours to days, many other Be stars seem to maintain their disks
in a relatively quiescent manner over several years (see, e.g.,
Dachs 1987; Telting 2000; Okazaki 2007).
It is worthwhile, then, to first consider a ``baseline'' disk
formation mechanism that feeds the disk with a more or less
steady-state supply of angular momentum from below.
Other proposed mechanisms then may act as sources for the
complex observed patterns of Be-star variability (see also
{\S}~6.2).

The outline of this paper is as follows.
{\S}~2 presents an overview of how stellar pulsations may be
coupled to the outward transport of mass and momentum in the
circumstellar regions of Be stars.
In {\S}~3 the detailed properties of waves and shocks are
presented, including how evanescent pulsation modes couple to
propagating waves, how these waves steepen into dissipative
shocks, and how the waves can exert a net pressure to affect
the time-steady properties of the stellar atmosphere.
{\S}~4 describes the time-independent conservation equations
for mass and momentum that are solved with contributions from
the waves, and {\S}~5 gives the results.
A discussion of the implications of these models on our overall
understanding of Be-star disk formation and time variability
is given in {\S}~6.
Finally, {\S}~7 contains a brief summary of the major results
and discusses how the proposed physical processes should be
further tested and refined.

\section{The Disk-Pulsation Connection}

Do stellar pulsations represent a causal factor in producing the
wide variety of observed manifestations of the Be phenomenon?
This possible connection has been discussed for several decades
both from the standpoints of Be-star observations (e.g.,
Baade 1982, 1985; Henrichs 1984; Willson 1986; Smith 1988) and
pulsation theory (Ando 1986; Osaki 1986; Castor 1986;
Saio 1994; Lee 2006, 2007).
For a while, there was significant debate over whether much of the
observed Be-star line profile variability was even the result of
NRP, or if these features were simply rotationally modulated
``spots'' (e.g., Baade \& Balona 1994).
However, recent improvements in the detection and modeling of NRP
in many Be stars has put the pulsational interpretation on firmer
footing (Rivinius et al.\  2003; Rivinius 2007; Townsend 2007).

The well-studied Be star $\mu$~Cen has been a key target in the search
for disk-pulsation correlations (e.g., Rivinius et al.\  1998).
The multiple NRP periods of $\mu$~Cen seem to undergo beating
in phase with outbursts of circumstellar material into the disk,
suggesting a direct input of mass and momentum when the pulsation
displacements are largest.
Increases in disk emission at NRP amplitude maxima have also been
observed for other multiperiodic Be stars such as $\zeta$ Oph
(Kambe et al.\  1993a) and 28 Cyg (Tubbesing et al.\  2000).
However, similar kinds of outbursts are also seen for Be stars
like $\omega$~CMa that seem to have only a single pulsation period
(\v{S}tefl et al.\  2003).
For other Be stars, there has been line profile variability
observed at Doppler velocities that {\em exceed the projected
photospheric rotation velocity} (Chen et al.\  1989;
Kambe et al.\  1993b).
This may be evidence for a radial increase in the rotation rate
between the photosphere and the inner edge of the disk.

In addition to observational connections between pulsations and
disks, there are many stars without disks (O, B, and Wolf-Rayet)
for which there exists coupled variability between the photosphere
and the stellar wind.
The radial pulsator BW Vul clearly shows nonlinear shock-like 
features in its photosphere (e.g., Smith \& Jeffery 2003) that are
in phase with radially accelerating features in its supersonic wind
(Massa 1994; Owocki \& Cranmer 2002).
Several stars show different periods for photospheric and wind
variability, with the former (due to NRP) often being shorter
than the latter (e.g., Howarth et al.\  1993, 1998;
Kaufer et al.\  2007).
Although much of the observed stellar wind variability may be
attributed to structures corotating with the star
(Kaper et al.\  1996), the distribution of these stars in luminosity
and effective temperature seems to agree roughly with the domain of
so-called ``strange-mode'' oscillations (Fullerton et al.\  1996),
indicating that there may be a pulsational cause for some types of
wind variability.

From a theoretical standpoint, it has been known for some time that
small-scale hydrodynamic fluctuations (waves, shocks, turbulence,
pulsations) can affect the large-scale mean properties of a fluid.
Linear waves propagating in an inhomogeneous medium exert a
time-steady {\em wave pressure} on the fluid even without dissipating
(Bretherton \& Garrett 1968; Dewar 1970; Jacques 1977).
Alfv\'{e}n wave pressure is believed to be a key contributor to
the acceleration of the high-speed solar wind (e.g., Cranmer 2004).
A nonlinear extension of this effect is likely to be acting in the
outer atmospheres of cool pulsating giants and supergiants---e.g.,
Mira variables and asymptotic giant branch stars---with pulsation
driven mass loss suspected to occur in many cases (Willson 2000;
Woitke 2007; Neilson \& Lester 2008).

In addition to supplying linear momentum, it has been shown that waves
can transport and deposit {\em angular momentum} in rotating systems.
Interior gravity waves in rotating planetary atmospheres have been
shown to be responsible for driving various kinds of mean shear flows
(Eliassen \& Palm 1960; Bretherton 1969; Plumb 1977; Schatzman 1993;
Rogers et al.\  2008).
In stellar interiors, the study of how diffusive transport processes
interact with the mean rotation profile has proceeded with the implicit
assumption that waves or turbulent eddies may be the cause of the
``anomalous'' transport (e.g., R\"{u}diger 1977; Zahn et al.\  1997;
Talon 2008).
Ando (1982) considered the coupling between waves and rotation as
a possible means of generating differential rotation in stellar
interiors.
Recently, Townsend \& MacDonald (2008) found that the interior
transport of angular momentum by pulsations may be an important
factor in the evolution of rotating massive stars.

The application of the above ideas to the episodic spinup and
mass loss of Be stars was developed by Ando (1986, 1988),
Osaki (1986, 1999), Lee \& Saio (1993), Saio (1994), and
Lee (2006, 2007).
This mechanism was originally believed to transport angular
momentum up to the stellar surface only for prograde NRP modes
(i.e., modes for which the azimuthal group velocity is in the same
direction as the mean rotation); see, e.g., Ando (1983).
A retrograde oscillation mode was thought to transport angular
momentum downward and thus potentially spin down the outer layers
of the star.
However, more recent work---which takes a more complete inventory of
the various second order wave-rotation coupling terms---finds that
some types of coupling do not depend at all on the sign of the
azimuthal group velocity (e.g., Lee 2007).
Furthermore, Townsend (2005) and Pantillon et al.\  (2007) showed
that some ``mixed modes'' may exist in rapidly rotating stars that
appear to have retrograde phase velocities but prograde group
velocities.
There are thus several potential ways for {\em retrograde} NRP modes
(which appear to be prevalent in Be stars) to be associated with
the transport of angular momentum up to the surface and beyond.

Aside from the work of Ando and colleagues, though, surprisingly
little work has been done to study how pulsational energy
(as well as the effects of pulsations on the mean fluid) may
``leak'' out of a hot star into its circumstellar medium.
Cranmer (1996) studied some aspects of how low-frequency evanescent
waves in B-star photospheres might evolve into propagating waves
at larger heights in a stellar wind.
Townsend (2000a,b,c) modeled NRP leakage from the perspective of
whether the $p$ and $g$ modes are propagating or evanescent in the
layers directly beneath the photosphere (see also Gautschy 1992).
The goal of this paper is to follow up on these ideas and to
investigate how an extended hot-star atmosphere responds to the
presence of pulsational oscillations at its photospheric base.

\section{Waves and Shocks}

This section contains a derivation of some of the physical processes
that determine how pulsationally driven waves can influence the mean
properties of a stellar atmosphere.
The subsections below isolate four of the main effects that
need to be considered: evanescence ({\S}~3.1),
resonant excitation at the acoustic cutoff ({\S}~3.2),
shock steepening and dissipation ({\S}~3.3),
and wave pressure ({\S}~3.4).
Some of the proposed connections between the above mechanisms
are still somewhat speculative, so it is hoped that this combination
of ideas will eventually be tested with numerical simulations.

This paper makes the implicit assumption that there is a reasonably
{\em time-steady} input stream of pulsational oscillations from
the stellar interior.
Observations suggest, however, that many NRP modes have finite
lifetimes, and that they undergo secular amplitude variations over
a wide range of time scales (see {\S}~6.2 for discussion of the
implications of NRP transience on Be-star variability).
The models presented below also ignore several effects that may
be important in some situations, such as
heat conduction, radiative damping, magnetic fields, and
ballistic freefall behind strong shocks (the latter being
important in, e.g., Mira variables).

\subsection{Evanescent Pulsation Modes}

For the purposes of modeling the effects of B-star NRP oscillations
on their outer atmospheres, it is important to first summarize what
is known about them observationally.
The $\beta$~Cep variables have typical pulsation periods of order
2--10 hrs, spectral types between B0 and B2, and are often
interpreted as low-order $p$ and $g$ modes
(Sterken \& Jerzykiewicz 1992; Aerts \& de Cat 2003).
Slowly pulsating B (SPB) stars have longer periods of 10--50 hrs,
later spectral types of B3 to B9, and they appear to be pulsating
in high-order $g$ modes (e.g., de Cat 2007).
Pulsating Be stars often have periods that appear similar to the SPB
class, but with some modes (including beat periods) that extend up
into the 100--200 hr range (e.g., Kaufer et al.\  2006).
Some B stars are ``hybrid'' pulsators that exhibit both $\beta$~Cep
and SPB type modes (Dziembowski \& Pamyatnykh 2008).

Figure 1 shows a collection of B-star pulsation frequencies
$\omega$ and horizontal wavenumbers $k_x$ that have been normalized,
respectively, by dividing by an estimated acoustic cutoff frequency
$\omega_a$ in the photosphere and multiplying by an estimated density
scale height $H$.
Spectral types, periods, and angular mode identifications were
obtained for the $\beta$~Cep and SPB stars from
de Cat (2002, 2003)\footnote{%
The 2005 version of this database was obtained from:
http://www.ster.kuleuven.ac.be/$\sim$peter/Bstars/}
and for Be stars from Rivinius et al.\  (2003).
Additional low-frequency modes for two hybrid B-type pulsators
(12~Lac and $\nu$~Eri) were taken from Dziembowski \& Pamyatnykh (2008),
and the assignment of $\ell = |m| = 2$ to these modes is relatively
uncertain.

\begin{figure}
\epsscale{1.13}
\plotone{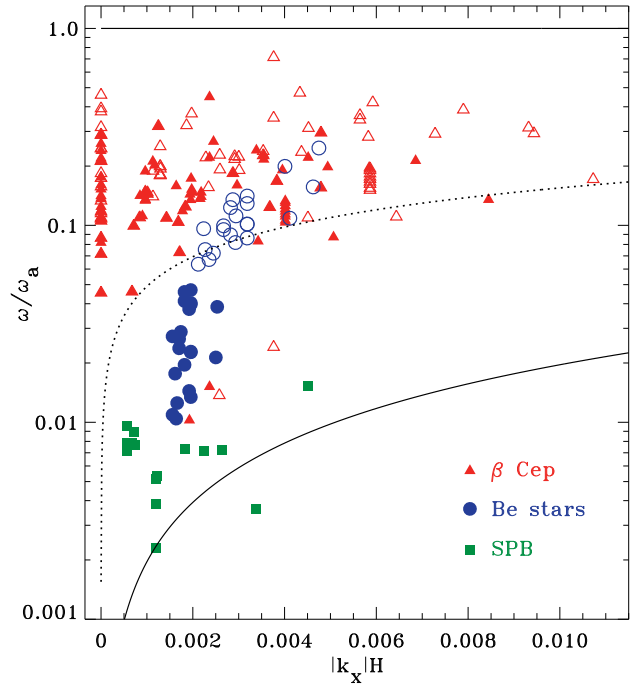}
\caption{Collected B-star pulsation frequencies and azimuthal
wavenumbers, normalized to dimensionless quantities via the
photospheric acoustic cutoff frequencies $\omega_a$ and scale
heights $H$.
Data for $\beta$~Cep variables ({\em red triangles}), SPB stars
({\em green squares}), and Be stars ({\em blue circles}) are
shown.
Filled symbols assume nonrotating stellar properties; open symbols
attempt to take account of rapid rotation.
Also shown are acoustic-gravity wave propagation boundaries
({\em solid lines}) and $f$-mode curve ({\em dotted line}).}
\end{figure}

The two-dimensional ($\omega, k_x$) plot illustrated by Figure 1 is
often called a diagnostic diagram for waves in stellar atmospheres.
The solid curves are the ``propagation boundary curves'' between
which acoustic-gravity waves are evanescent.
Above the upper curve, waves propagate vertically as acoustic
or $p$-mode waves; below the lower curve, they propagate as
internal gravity or $g$-mode waves  (e.g., Lamb 1908, 1932;
Mihalas \& Mihalas 1984).
To determine the normalizing values of the acoustic cutoff frequency
$\omega_a$ and scale height $H$ for each star, the tabulated spectral
types and luminosity classes were converted into stellar masses
$M_{\ast}$, radii $R_{\ast}$, and effective temperatures
$T_{\rm eff}$ in a single uniform way; i.e., using the luminosity
and $T_{\rm eff}$ relations of de Jager \& Nieuwenhuijzen (1987),
and interpolating onto the evolutionary tracks of Claret (2004)
to obtain masses (see also {\S}~2 of Cranmer 2005).
The magnitude of the horizontal wavenumber was computed from the
tabulated meridional degree number $\ell$ and the assumption that
$\ell = |m|$, the latter being the azimuthal mode number.
Thus, $|k_{x}| = |m| / R_{\ast}$.
For the Be stars, only the modes identified as having $m=2$ by
Rivinius et al.\  (2003) are shown.

The solid symbols in Figure 1 show values computed using the
standard expressions for $\omega_a$ and $H$ for nonrotating stars
(see below).
The open symbols show modified values that take account of two
attempts at improving the accuracy of these normalizations:
\begin{enumerate}
\item
The surface gravity was reduced by taking account of both
centrifugal and radiative acceleration.
A lower limit to the stellar rotation was applied by assuming that
the observed projected rotation rate $v \sin i$ is equal to the
actual equatorial rotation speed $V_{\rm rot}$.
For the Be stars, it was assumed that the rotation speed is never
smaller than $V_{\rm crit}/2$ (see Cranmer 2005).
The radiative acceleration in the photosphere was estimated using
a fit to the numerical results of Lanz \& Hubeny (2007) given by
equations (\ref{eq:gedd})--(\ref{eq:grad}) below.
\item
For the Be stars only, the observed pulsation frequency was replaced
by an estimate of the intrinsic pulsation frequency in the rotating
frame, assuming that the Be-star oscillations are retrograde
(Rivinius et al.\  2003).
In this case, the rotating-frame frequency $\omega$ is higher
than the observed inertial-frame frequency $\omega_0$, with
\begin{equation}
  \omega \, \approx \, \omega_{0} + m\Omega (1 - C_{n\ell})
\end{equation}
and $C_{n\ell} \approx 0.2$ for low-order $f$ and $g$ modes
(e.g., Ledoux 1951).
We ignored second order centrifugal terms in the above expression
(i.e., terms proportional to $\Omega^2 / \omega$) since for
low-order $g$ modes with $\ell = |m| = 2$ they would give a
negligibly small correction (Saio 1981).
\end{enumerate}
For the SPB stars, these corrections were found to be unimportant
and are not shown.

Taking account of the above effects results in smaller values of
$\omega_a$ and in larger values of both $\omega$ and $H$, thus
moving each data point upwards and to the right in Figure 1.
Even with these corrections, though, nearly all of the observed
pulsation modes are still solidly evanescent (i.e., not
propagating vertically) in the photosphere.
Also, the Be-star data points seem to have moved from the
upper edge of the SPB region into the $\beta$~Cep region.
(Many of them seem to cluster around the fundamental-mode
curve, given by $\omega^{2} = g k_{x}$, which is a rough boundary
between the $p$-mode and $g$-mode regions.)
This may suggest that classical Be-star pulsations have a similar
intrinsic character as those of the $\beta$~Cep variables, despite
the apparent similarity of their periods to those of the SPB stars
(see also Baade 1982).

The observed ranges of oscillation parameters can be put into context
by using a theoretical model of {\em acoustic-gravity waves.}
The model presented here makes the simplifying assumptions of
an isothermal (constant temperature $T$) medium, plane-parallel
geometry, and a constant gravitational acceleration $g$.
It is also straightforward to first consider the system in the
rotating frame, in which we define the frequency $\omega$.
In this atmosphere, the sound speed is constant and is given by
\begin{equation}
  c_{s} \, = \, \sqrt{\frac{\gamma P_0}{\rho_0}} \, = \,
  \sqrt{\frac{\gamma k_{\rm B} T}{\mu m_{\rm H}}}  \,\, ,
\end{equation}
where $k_{\rm B}$ is Boltzmann's constant, $m_{\rm H}$ is the mass
of a hydrogen atom, and $\mu$ is the mean molecular weight of the gas.
The latter quantity is assumed to be $\mu = 0.604$, which corresponds
to a standard (10\% helium) elemental abundance mixture.
The adiabatic exponent $\gamma$ is the ratio of specific heats
$c_{p}/c_{v}$, and usually can range between 1 (isothermal) and 5/3
(adiabatic).
For an isothermal atmosphere, the pressure and density scale heights
are equal to one another and given by $H = c_{s}^{2}/ \gamma g$
(which does not depend on the value of $\gamma$ itself).

The standard assumption in linear wave theory is to model the
physical variables (i.e., density, velocity, pressure) as the
sum of a time-steady (zeroth order) component and an oscillating
(first order) component.
The linear amplitudes vary as $e^{i\omega t -ik_{z}z -ik_{x}x}$
with a real frequency $\omega$.
We model the conditions in the equatorial plane, with gravity acting
in the $-z$ direction and the stellar rotation proceeding in the
$+x$ direction (see Appendix).
The azimuthal wavenumber can thus be expressed in conventional
NRP terminology as $k_{x} = -m / R_{\ast}$, with negative [positive]
values of $m$ corresponding to prograde [retrograde] modes.
The time-steady vertical velocity $w_0$ is assumed to be zero, and the
azimuthal velocity $u_0$ is also set to zero (in the rotating frame).
The zeroth-order density $\rho_{0}$ is proportional to $e^{-z/H}$.
The corresponding linear amplitudes are denoted $w_1$, $u_1$, and
$\rho_1$, and they are modeled as complex quantities in order to
account for phase differences between their respective oscillations.

For a given frequency $\omega$ and horizontal wavenumber $k_x$,
the vertical wavenumber is in general a complex quantity given by
\begin{equation}
  k_{z} \, = \, \frac{i}{2H} \pm \left[
  \frac{(\omega^{2} - \omega_{a}^{2})}{c_{s}^2} -
  \frac{(\omega^{2} - \omega_{g}^{2}) \, k_{x}^{2}}{\omega^2}
  \right]^{1/2}  \,\, ,
\end{equation}
where the acoustic cutoff frequency is $\omega_{a}= \gamma g / 2c_{s}$
and the Brunt-V\"{a}is\"{a}l\"{a} frequency is
$\omega_{g} = (\gamma - 1)^{1/2} g/c_{s}$ (see, e.g.,
Lamb 1932; Mihalas \& Mihalas 1984; Wang et al.\  1995).
The regions of $(\omega, k_{x})$ space where the vertical wavenumber
has a real part correspond to propagating waves, and the regions where
$k_z$ is completely imaginary correspond to evanescence.
If the amplitude and phase of the vertical velocity amplitude $w_1$
are specified, the horizontal velocity and density fluctuations are
given in dimensionless form by
\begin{equation}
  \frac{u_1}{w_1} \, = \,
  \frac{c_{s}^{2} k_x}{\omega^{2} - c_{s}^{2} k_{x}^2}
  \left( k_{z} - \frac{i}{\gamma H} \right)
\end{equation}
\begin{equation}
  \frac{\rho_{1} / \rho_{0}}{w_{1} / c_s} \, = \,
  \frac{\omega c_s}{\omega^{2} - c_{s}^{2} k_{x}^2}
  \left\{ k_{z} + \frac{i}{H} \left[
  \frac{(\gamma - 1) c_{s}^{2} k_{x}^2}{\gamma \omega^2} - 1
  \right] \right\}  \,\, .
\end{equation}
Note that when $k_z$ is completely imaginary, the above expressions
are also imaginary.
This implies that in the case of evanescent waves, $u_1$ and $w_1$ are
{90\arcdeg} out of phase, and $\rho_1$ and $w_1$ are also {90\arcdeg}
out of phase.

It is useful to illustrate how these waves behave in the limit of
small horizontal wavenumbers (i.e., $|k_{x} H| \ll 1$) which applies
to B-star NRP observed at the stellar surface (see Figure 1).
It is also useful to use the simplifying limit of $\gamma = 1$, which
corresponds to isothermal fluctuations.
This approximation is consistent with the existence of rapid radiative
heating and cooling in the relatively dense outer atmospheres of hot
O and B stars (e.g., Drew 1989; Millar \& Marlborough 1999;
Carciofi \& Bjorkman 2006).
With the above assumptions, the lower propagation boundary disappears
and the upper propagation boundary is greatly simplified: i.e.,
waves with $\omega > \omega_a$ propagate; waves with
$\omega < \omega_a$ are evanescent.

Propagating waves have exponentially growing amplitudes with height,
with the non-oscillating part of the height dependence being
given by $w_{1} \propto e^{z/2H}$.
These waves have a vertical phase speed
\begin{equation}
  V_{\rm ph} \, = \, \frac{\omega}{\mbox{Re} \, k_z}
  \, \approx \, c_{s} \left( 1 -
  \frac{\omega_{a}^2}{\omega^2} \right)^{-1/2}
\end{equation}
and a group velocity $V_{\rm gr} = c_{s}^{2} / V_{\rm ph}$.
For upwardly propagating waves, the above amplitude relations 
can be simplified to show that
\begin{equation}
  \mbox{Re} \left( \frac{u_1}{w_1} \right) \, \approx \,
  2 k_{x} H \, \frac{\omega_a}{\omega} \,
  \frac{c_s}{V_{\rm ph}} \, \ll 1
  \label{eq:u1w1simp}
\end{equation}
\begin{equation}
  \mbox{Re} \left( \frac{\rho_{1} / \rho_{0}}{w_{1} / c_s}
  \right) \, \approx \, \frac{c_s}{V_{\rm ph}} \,\, .
\end{equation}
Note that the sign of $k_x$ determines the sign of the magnitude
ratio $u_{1}/w_{1}$, but the ratio $\rho_{1}/w_{1}$ is positive
definite.

Evanescent waves have two possible solutions for their (completely
non-oscillatory) height dependence; one ``steeper'' than the
propagating solution, and one ``shallower.''
The shallow [steep] solution has a total wave energy density that
decays [grows] with increasing height.
Thus, the shallow solution is the physically realistic choice when
considering waves that originate at low heights and have effects
on the medium at larger heights (see, e.g., Wang et al.\  1995).
The vertical amplitude dependence of the shallow solution is given by
\begin{equation}
  w_{1} \, \propto \, \exp \left[ \frac{z}{2H} \left( 1 -
  \sqrt{1 - \frac{\omega^2}{\omega_{a}^2}} \right) \right]  \,\, .
  \label{eq:w1ev}
\end{equation}
For evanescent waves, the phase speed is formally infinite and
the group velocity is zero.
However, Cranmer (1996) showed that if a time-steady
{\em subsonic wind} ($w_{0} \ll c_{s}$) is included in the linear
equations for acoustic-gravity waves, the shallow evanescent solution
can be shown to have a large---but finite---upward phase speed of
order $c_{s}^{2}/w_{0}$.
The steep evanescent solution corresponds to a downward phase
speed of $-c_{s}^{2}/3w_{0}$.
These both correspond to negligibly small (subsonic) group
velocities and do not need to be considered further.

\subsection{Resonant Wave Excitation}

It has been known for some time that the acoustic cutoff $\omega_a$
is a preferred resonance frequency for an atmosphere that has been
disturbed by a pulse or piston-like initial condition (see, e.g.,
Lamb 1908; Schmidt \& Zirker 1963).
The acoustic cutoff frequency also acts as a fundamental ``ringing''
mode for disturbances of a sinusoidal nature (Fleck \& Schmitz 1991;
Kalkofen et al.\  1994; Sutmann et al.\  1998).
Of particular interest here is the case of an evanescent oscillation
driven by NRP (with $\omega < \omega_a$) that can excite a resonant
oscillation ($\omega \approx \omega_a$).\footnote{%
In addition to long-period evanescent waves giving rise to resonant
excitation, it is important to note also that {\em shorter} period
acoustic waves may also lead to resonant excitation at the cutoff,
possibly via nonlinear effects such as ``shock cannibalization''
(see Rammacher \& Ulmschneider 1992).  Thus, there appear to be
several independent ways to create waves at the cutoff frequency.}

The one-dimensional problem of resonant excitation in an infinite
isothermal atmosphere can be solved analytically (e.g.,
Fleck \& Schmitz 1991).
In this section we make use of this simple model to gain insight
about the physics of resonant waves.
Although the model contains features that are formally inapplicable
to actual pulsating stars (i.e., the assumption of an instantaneous
``start'' at $t=0$), it remains valuable as a way to obtain
numerical estimates of the resonant wave properties in the absence
of a more accurate theory.
Let us then consider a piston with an average height slightly below
the photosphere ($z_{\rm piston} < 0$) with a vertical pulsation
amplitude $V_{\rm puls}$.
For simplicity, the long-wavelength horizontal velocity
fluctuations can be ignored; i.e., $k_{x} = 0$.
Thus, for a piston oscillating with evanescent frequency $\omega$,
the height dependence of this driven mode is given by
equation (\ref{eq:w1ev}), and its time dependence at larger heights
remains strictly sinusoidal.
In addition, when the piston motion begins (at time $t=0$) there
arises a resonant response at the acoustic cutoff frequency.
These two components of the vertical velocity amplitude can be
expressed as
\begin{displaymath}
  w_{1} (z,t) \, = \, V_{\rm puls} e^{-i \omega t + \kappa Z}
  \,\, +
\end{displaymath}
\begin{equation}
  V_{\rm res}(z,t) \left[
  \sin \left( \omega_{a} t - \frac{3\pi}{4} \right) +
  \frac{i\omega}{\omega_{a}}
  \cos \left( \omega_{a} t - \frac{3\pi}{4} \right) \right]
\end{equation}
where $Z \equiv (z - z_{\rm piston})/2H$ and
$\kappa = 1 - (1 - \omega^{2}/\omega_{a}^{2})^{1/2}$
(Kalkofen et al.\  1994; Sutmann et al.\  1998).
In the late-time limit of $t \gg z /c_{s}$, after which the initial
transient disturbance has propagated away, the resonant component's
velocity amplitude is given by
\begin{equation}
  V_{\rm res} (z,t) \, = \, V_{\rm puls}
  \sqrt{\frac{2 \omega_a}{\pi}} \,
  \frac{Z e^{Z}}{(\omega_{a}^{2} - \omega^{2}) \, t^{3/2}}  \,\, .
  \label{eq:Vfree}
\end{equation}
This expression gives only the lowest order term in an infinite
expansion; the higher order terms have more rapid time decay (e.g.,
$t^{-5/2}$, $t^{-7/2}$, etc.) and thus the above term dominates the
late-time ($t \gg z /c_{s}$) behavior of the resonant mode.

For a given time $t$ after the piston begins to oscillate, both
the driven (evanescent) mode and the resonant (cutoff) mode coexist
in the atmosphere.
Immediately above the piston, $V_{\rm res} \ll V_{\rm puls}$,
but as one proceeds higher in the atmosphere, the radial growth of
the resonant mode overtakes that of the shallow evanescent mode
(eq.~[\ref{eq:w1ev}]) and the resonant mode becomes stronger.
It is suspected that the stellar photosphere sits {\em below}
the height where this transition occurs, because the observable
signatures of NRP show only the evanescent driver.
A major conjecture of this paper is that the upper layers of the
atmosphere---including the ``inner edge'' of the Keplerian
disk---are those where the resonant oscillations are dominant.

Equation (\ref{eq:Vfree}) shows that the resonant oscillation
amplitude decays in time as $t^{-3/2}$.
If left undisturbed it would decay eventually to zero, leaving
only the driven evanescent mode.
However, this property of the solution seems to be an artifact of
the idealized nature of an infinitely extended, isothermal, and
plane-parallel atmosphere.
For a {\em real} atmosphere, there are several effects that have
been suggested to arrest this decay and produce a quasi-steady
stream of resonant waves:
\begin{enumerate}
\item
Fleck \& Schmitz (1991) found that a sharp reflecting boundary at some
height above the piston (even a relatively permeable one) can cause
enough downward reflection of the resonant waves to maintain them
at a finite amplitude over long times.
Reflection effectively ``restarts the clock'' on the time $t$ in
equation (\ref{eq:Vfree}) every time the piston sees a reflected wave.
\item
Radial gradients in the sound speed or scale height give
rise to {\em gradual} linear reflection that can lead to continual
excitation of the resonant waves (e.g., Pitteway \& Hines 1965;
Schmitz \& Fleck 1992; Lou 1995; Musielak et al.\  2005;
Erd\'{e}lyi et al.\  2007; Taroyan \& Erd\'{e}lyi 2008).
This kind of reflection is also believed to be acting on Alfv\'{e}n
waves in the solar chromosphere and corona, and in that case it may be
responsible for seeding an ongoing turbulent cascade
(Heinemann \& Olbert 1980; Matthaeus et al.\  1999;
Cranmer \& van Ballegooijen 2005).
\item
If the driven evanescent oscillations are multiperiodic or
stochastic in nature, the resulting departures from simple sinusoidal
piston motion could result in continuously excited resonant waves.
Such intermittency could be the result of rotationally generated
subsurface convection (Espinosa Lara \& Rieutord 2007;
Maeder et al.\  2008), NRP mode beating (Rivinius et al.\  1998),
or other nonlinear processes that can cause one NRP mode to grow
at the expense of another mode (see, e.g., Smith 1986).
\end{enumerate}

The remainder of this paper assumes the existence of a continual
regeneration of resonant oscillations in B-star atmospheres.
The general phenomenon of wave reflection seems to be the most
likely route for this to occur.
Fleck \& Schmitz (1991) showed how the presence of wave reflection
may cause the resonant waves to ``forget'' about an instantaneous
initial condition and become essentially self-sustaining. 
Although the reflection in the models of Fleck \& Schmitz (1991) was
induced by a sharp upper boundary, there are several other ways
that reflection can be produced gradually in an actual stellar
atmosphere (see the second item in the above list).
In the absence of a robust theory or simulation of these effects
in hot-star atmospheres, however, we are limited to using simpler
prescriptions such as equation~(\ref{eq:Vfree}).

The use of the simple theory derived above should not be
interpreted as a suggestion that resonant waves are produced by
an abrupt initial condition.
This model is used only to estimate the mean amplitude of a
steady-state ensemble of resonant waves, by choosing a
representative time $t$ in equation (\ref{eq:Vfree}).
Equivalently, this can be specified in terms of the height
$\Lambda$ of a fictional reflecting boundary above the photosphere,
and the reflection time is thus estimated as $t \approx \Lambda / c_s$
(see, e.g., Figure 2 of Fleck \& Schmitz 1991).
The parameter $\Lambda$ is used as a free parameter in the
models discussed below.
Eventually, of course, there should be time-dependent numerical
simulations of the partial wave reflection and resonant wave
excitation in B-star atmospheres that would allow the most
appropriate values of $\Lambda$ to be determined.

What values of $\Lambda$ should be considered realistic?
It is useful to estimate the heights in the atmosphere at which
the fluid parameters are thought to undergo qualitative changes.
These are the heights at which substantial wave reflection should
occur.
One suggestion is to compute the height at which the strongest wave
modes steepen into shocks (and probably saturate in amplitude).
This onset of nonlinearity signals a rapid change in the overall
properties of the atmosphere.
For hot stars with NRP amplitudes that are already of the same order
as the sound speed {\em in the photosphere,} the height at which
they steepen into shocks may be as low as 2 to 4 photospheric
scale heights above the photosphere itself.
For a star with an accelerating wind, other significant heights 
would be the sonic and critical points (which are not the same
for a radiatively driven wind; see Castor et al.\  1975).
For the standard B2~V star model discussed in detail in {\S}~5, the
sonic point would be at a radius of about 10 scale heights.
The critical point is another scale height or two above the
sonic point.

It is possible, though, that the relevant time scale $t$ could be
much larger than suggested above.
For a B star, a value of $\Lambda \approx 10 H$ corresponds to
a time that is only about 10\% of a single evanescent NRP period.
If $t$ is determined instead by long-term secular changes in
the NRP ``piston'' (i.e., beating or mode growth/decay), it may be
as large as the mode lifetimes themselves; typically no smaller than
10 to 100 NRP periods.
However, since there is also evidence for shorter time scales
being important (consistent with ongoing wave reflection from the
upper atmosphere), we will make the order-of-magnitude assumption
that $\Lambda$ is of order 10 scale heights and vary it up
and down from that baseline value.

Figure 2 compares the height dependences of vertical velocity
amplitudes for evanescent and resonant waves.
The stellar parameters assumed here are those of the baseline B2~V
model discussed in detail in {\S}~5.
The evanescent NRP driver has a period of 10 hrs, the acoustic cutoff
period is 1.3 hrs, and the piston is assumed to sit $0.5 H$ below
the photosphere.
For an assumed value of $\Lambda = 10 H$, the resonant mode becomes
stronger than the evanescent mode at a height of approximately 4
scale heights above the photosphere (i.e., when the hydrostatic
density has dropped by a factor of $\sim$50 from the photospheric value).
Also shown is an upper-limit velocity amplitude $V_{\rm max}$ that
illustrates the total energy density available for waves in a
stratified atmosphere.
This quantity was computed under the assumption that
$\rho_{0} V_{\rm max}^{2}$ is constant, and $V_{\rm max}$ is
normalized by the evanescent NRP driving amplitude at the piston.
The thick curves show linear undamped solutions (i.e.,
eq.~[\ref{eq:Vfree}]) and the thin curves show the result of
applying the shock steepening, dissipation, and wave pressure
effects discussed below.

\begin{figure}
\epsscale{1.11}
\plotone{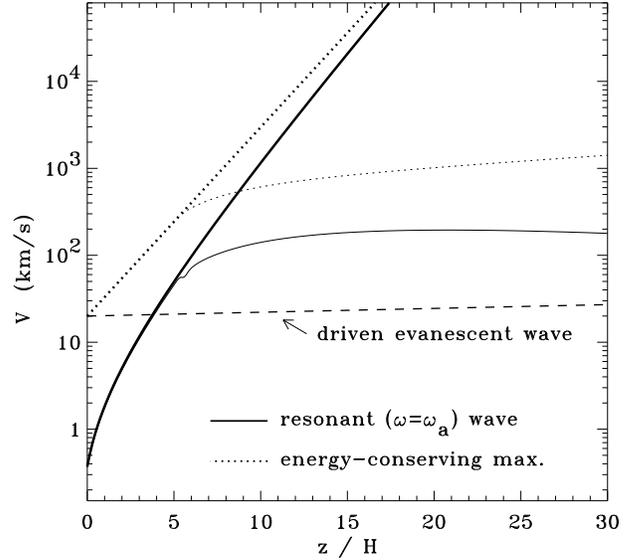}
\caption{Comparison of height dependences for driven evanescent wave
amplitudes $V_{\rm puls}$ ({\em dashed line}), resonantly excited
wave amplitudes $V_{\rm res}$ ({\em solid lines}), and the maximum
available wave amplitudes $V_{\rm max}$ from energy conservation
({\em dotted lines}).
Thick lines denote ideal linear results; thin lines show numerical
results that take account of wave dissipation and shock steepening.}
\end{figure}

Note that the radial growth of the undamped resonant mode is more
rapid than for a classical acoustic-gravity wave at
$\omega \geq \omega_a$ (the latter illustrated by
$V_{\rm max} \propto e^{Z}$).
The extra linear factor of $Z$ in equation (\ref{eq:Vfree})
is a unique feature of the resonantly excited oscillation.
This factor has the strongest impact very near the piston height,
but grows progressively less important once one goes a few scale
heights above the piston.
The resulting properties of the upper atmosphere are thus relatively
insensitive to the exact value chosen for $z_{\rm piston}$.
Also, when wave damping and wave pressure are taken into account
(i.e., the thin curves in Fig.\  2) the resonant mode no longer
grows faster than $V_{\rm max}$ at all.

Choosing larger values of the reflection height $\Lambda$
results in a lower ``efficiency'' in the conversion of driving
wave energy to resonant wave energy.
This efficiency is estimated as the ratio $V_{\rm res}/V_{\rm max}$,
and Figure 3 shows how this varies as a function of the $\Lambda$
parameter itself.
The models that were used to compute these efficiencies
contained the full range of shock steepening and wave pressure
effects (see {\S\S}~3.3--3.4), and they constrained $V_{\rm res}$ to
never exceed the maximum available energy specified by $V_{\rm max}$.
The efficiencies are shown both at the photospheric base and
at the height of peak efficiency.
The latter occurs usually a few scale heights above the photosphere,
before substantial damping sets in.
It is clear that values of $\Lambda/H$ less than about 3 or 4
may not be physically realistic, since they result in the resonant
wave attempting to extract more energy from the driving mode
than is actually available.

\begin{figure}
\epsscale{1.11}
\plotone{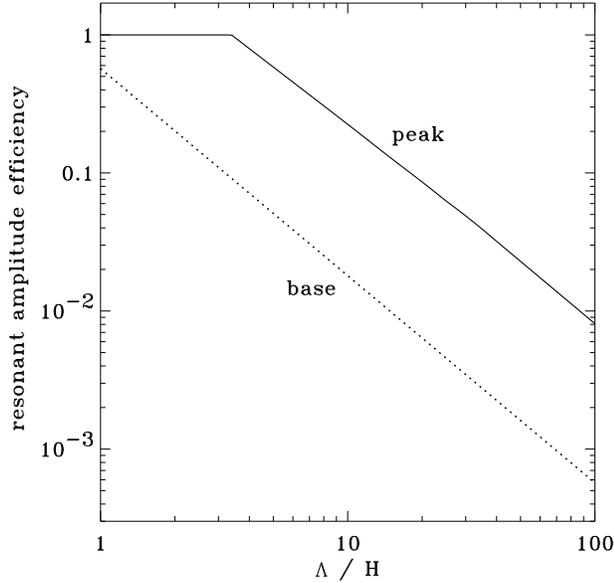}
\caption{Efficiency of resonant excitation ($V_{\rm res}/V_{\rm max}$)
plotted as a function of the assumed reflection height $\Lambda$
for a series of B2~V atmosphere models (see text).
The photospheric ({\em dotted line}) and maximum ({\em solid line})
values for each model are shown.}
\end{figure}

At the subphotospheric height of the idealized piston, the models
below ignore the effects of radiative acceleration $g_{\rm rad}$
(which can counteract gravity at and above the photosphere).
Lanz \& Hubeny (2007) reported the existence of a local maximum in
$g_{\rm rad}$ at the $\tau \approx 1$ photosphere, with less of an
effect in the deeper layers that have smaller temperature gradients.
Thus, because the net gravitational acceleration is higher at the
piston than in the photosphere, the acoustic cutoff frequency at the
piston is assumed to be slightly higher than that at the photosphere.
For the standard B2~V star model, this results in a resonantly
excited wave having a frequency about 1.2 times the {\em local}
acoustic cutoff frequency once it reaches the photosphere.
In other words, the assumption is that the wave eventually
``forgets'' about its origin as a resonant excitation and just
propagates (as any other linear wave would) with a constant
frequency.

\subsection{Wave Action Conservation and Shock Dissipation}

The resonantly excited wave modes discussed above
($\omega \gtrsim \omega_a$) are assumed to propagate upward from
the photosphere and grow in amplitude.
At some point, these mainly longitudinal fluctuations are expected
to steepen into shocks and begin to dissipate some of their energy
(e.g., Castor 1986).
The nonlinear evolution of such a wave train is modeled here using a
formalism described more completely by Cranmer et al.\  (2007).
The wave energy density $U_s$ is assumed to obey an equation of
wave action conservation,
\begin{equation}
  \frac{\partial}{\partial t} \left( \frac{U_s}{\omega'}
  \right) + \frac{1}{A} \frac{\partial}{\partial z}
  \left[ \frac{(w_{0} + V_{\rm gr}) A U_s}{\omega'} \right] \, = \,
  - \frac{Q_{\rm sh}}{\omega'}
  \label{eq:dUsdt}
\end{equation}
(see also Jacques 1977; Koninx 1992), where $A$ is the cross-sectional
area of a flow tube and $Q_{\rm sh}$ is the shock dissipation rate.
The time derivative term in equation (\ref{eq:dUsdt}) is given only
for completeness and is ignored in the time-stationary models below.
In general, the Doppler shifted frequency in the moving frame
(i.e., taking account of both rotation and a stellar wind) is given
by $\omega' = \omega - u_{0} k_{x} -  w_{0} k_{z}$.
As above, the $u_0$ effect is ignored by interpreting $\omega$ as the
corotating frequency, and the factor dependent on $w_0$ is taken into
account by using the proper dispersion relation.
The wave energy density is defined as
\begin{equation}
  U_{s} \, = \, \frac{1}{s} \rho w_{1}^{2}
  \label{eq:Usdef}
\end{equation}
where $s$ is a dimensionless shape factor determined by the spatial
profile of the waves.
The peak-to-peak vertical velocity amplitude is denoted $w_{1}$,
and the variance $w_{1}^{2}$ is equivalent to the magnitude of the
product $w_{1}^{\ast} w_{1}$ (see Appendix).

The adopted area function $A(z)$ contains both the spherical
expansion of the stellar atmosphere and an additional linear factor
(depending on the distance above the piston) that describes how the
resonantly excited waves differ from standard acoustic-gravity waves
(eq.~[\ref{eq:Vfree}]).  Thus,
\begin{equation}
  A \, \propto \, \frac{r^2}{(z - z_{\rm piston})^2}
\end{equation}
where $r$ is measured from the center of the star, and thus only
varies by a few percent over the heights considered here.
The models use the assumption that $z_{\rm piston}$ is $0.5 H$
below the photosphere.
Although $A$ diverges unphysically at $z = z_{\rm piston}$, this
location is always kept outside the modeled grid of heights.
The precise normalization of $A$ is unimportant, and only its
relative variation with height enters into equations (\ref{eq:dUsdt})
and (\ref{eq:wafull}).

The Doppler shifted frequency $\omega'$ can be computed in terms of the
comoving-frame phase speed $V_{\rm ph}$, and a more useful version of the
wave action conservation equation can be written as
\begin{equation}
  \frac{\partial}{\partial r} \left[
  \frac{(w_{0} + V_{\rm gr})(w_{0} + V_{\rm ph}) A U_{s}}{V_{\rm ph}}
  \right] \, = \, - \frac{(w_{0} + V_{\rm ph}) A Q_{\rm sh}}{V_{\rm ph}}
  \,\, .
  \label{eq:wafull}
\end{equation}
At the photospheric base, the lower boundary condition on the
wave energy density is specified as $U_{s} = \rho V_{\rm res}^{2}/2$,
with $s=2$ (i.e., a sinusoidal wave shape) also assumed at this height.

Before proceeding to describe the shock steepening in more detail,
it is useful to examine the wave action equation in a simpler
limiting case.
For a negligibly small vertical wind speed $w_0$, and for waves
sufficiently above the acoustic cutoff frequency that
$V_{\rm ph} \approx V_{\rm gr} \approx c_{s}$,
equation (\ref{eq:wafull}) simplifies greatly to become
\begin{equation}
  c_{s} \frac{\partial U_s}{\partial r} \, = \, -Q_{\rm sh} \,\, ,
  \label{eq:wasimp}
\end{equation}
where the additional assumption of $A \approx \mbox{constant}$ was
also applied for large heights $z \gg z_{\rm piston}$ in a
plane-parallel atmosphere.
This equation has the benefit of clearly illustrating how shock
dissipation affects the radial gradient of the wave energy density.
The other terms that have been stripped away are {\em non-dissipative}
contributors to the wave energy gradient.
Although the numerical solutions presented in {\S}~5 utilize the
more complete equation of wave action conservation
(eq.~[\ref{eq:wafull}]), the simplified form of equation
(\ref{eq:wasimp}) is later found to be useful to derive a
physically meaningful expression for the wave pressure.

A standard way of expressing the time-averaged shock dissipation rate is
\begin{equation}
  Q_{\rm sh} \, = \, \frac{\rho T \Delta S}{2\pi / \omega} \,\, ,
  \label{eq:Qs}
\end{equation}
where $\Delta S$ is the net entropy jump across a shock, and it is
nonzero only above the height where the wave train has steepened
into shocks.
This expression uses an approximation from so-called
``weak-shock theory'' that the volumetric heating rate is given
by the internal energy dissipated at one shock divided by the
mean time between shock passages in a periodic train.
This assumption breaks down for very strong radiative shock trains
(e.g., Carlsson \& Stein 1992, 1997), which dissipate their energy
in relatively narrow zones behind each shock.
However, the models presented below do not develop such strong shocks.
The gain in internal energy across an ideal inviscid shock is given by
\begin{equation}
  T\Delta S \, = \, c_{v} \left[ T_{2} - T_{1} \left(
  \rho_{2} / \rho_{1} \right)^{\gamma-1} \right]
  \label{eq:TdS1}
\end{equation}
where $c_v$ is the specific heat at constant volume for an ideal gas,
and subscripts 1 and 2 denote quantities measured on the
upstream (supersonic) and downstream (subsonic) sides of the shock
(Landau \& Lifshitz 1959).
The above expression is inconvenient for the limit of isothermal
shocks ($\gamma=1$) since the term in square brackets approaches
zero and the specific heat $c_v$ is proportional to $(\gamma-1)^{-1}$
and thus diverges to infinity.
Their product remains finite, and can be rewritten in this case as
\begin{equation}
  T\Delta S \, = \, c_{s}^{2}
  \left[ \left( \frac{M_{1}^{4}-1}{2 M_{1}^2} \right) -
  \ln M_{1}^{2} \right]
  \label{eq:TdS2}
\end{equation}
where $M_1$ is the Mach number of the shock.
The above expressions are valid for shocks of arbitrary strength and
are {\em not} limited to the traditional weak-shock approximation.

In order to use the shock dissipation rate in the wave action
conservation equation, the Mach number $M_1$ needs to be expressed in
terms of the local velocity amplitude $w_1$.
Cranmer et al.\  (2007) described a method of following the steepening
of an initially sinusoidal wave profile ($s=2$) into a fully steepened
sawtooth or N-wave ($s=3$).
Those steepening equations are modified here slightly due to the fact
that $V_{\rm ph} \neq c_{s}$ for acoustic-gravity waves having vertical
wavelength $\lambda = 2\pi V_{\rm ph}/ \omega$.
The instantaneous distance between a wave crest and the zero-velocity
node immediately ahead of it is given by
\begin{equation}
  \Delta z \, = \, \frac{\lambda_0}{4} - \frac{\gamma + 1}{2}
  \int \frac{w_{1} \, dz}{V_{\rm ph}} \,\, ,
  \label{eq:Delz}
\end{equation}
and the integration is taken from the photospheric lower boundary
(at which $\lambda = \lambda_0$) up to a given height.
The progressive nonlinear steepening of a wave gives rise to a
{\em decrease} in $\Delta z$ as the wave train propagates upwards.
At every point above the photosphere, the local steepening factor
$\zeta \equiv \Delta z / \lambda$ is known and is used to compute
both the shape factor $s$ and a shock efficiency
$\varepsilon = (M_{1}-1) c_{s} / w_{1}$.
The latter quantity is the ratio of the shock velocity amplitude
to the full velocity amplitude of the wave profile, and it can range
between zero (for a sinusoidal wave that has not yet steepened) and
one (once the crest overtakes the node ahead of it and $\zeta \leq 0$).
Cranmer et al.\  (2007) gave both numerical results and analytic
fitting formulae for $s(\zeta)$ and $\varepsilon(\zeta)$.
The models presented in {\S}~5 make use of these analytic fits.

\subsection{Wave Pressure}

Waves that propagate up from the stellar photosphere, steepen into
shocks, and dissipate a fraction of their energy also are able to
transfer a mean momentum flux (i.e., exert a ponderomotive force)
to the bulk atmosphere.
Although in some situations it is possible for dissipationless waves
to exert pressure (or Reynolds stresses) in a fluid, the ability for
{\em dissipating} waves to do so is much easier to understand
(e.g., Goldreich \& Nicholson 1989; Koninx 1992).
The oscillatory energy lost via dissipation is sufficiently ``randomized'' 
to produce a net, time-averaged source of both linear and angular
momentum.

In the Appendix, the second-order wave pressure contributions
to the radial and azimuthal momentum conservation equations are
computed from first principles.
This derivation does not make use of the standard linearizing
assumption that the first-order oscillations are smaller in
magnitude than the corresponding time-steady (zeroth order)
equilibrium properties.
It does, however, assume that the variations of the first-order
quantities in time ($t$) and in the azimuthal direction ($\phi$
or $x$) are sinusoidal.
The resulting momentum source terms are given in equations
(\ref{eq:Sxfin}) and (\ref{eq:Szfin}), and the wave-pressure
acceleration is thus expressed as
\begin{equation}
  {\bf g}_{\rm wp} \, = \, -\langle {\bf S} \rangle / \rho_{0}
\end{equation}
where the vector ${\bf g}_{\rm wp}$ is defined as positive on the
right-hand side of the radial momentum conservation equation
(eq.~[\ref{eq:wcons}]) when it points away from the star.

The radial component of the wave-pressure acceleration is given by
applying equation (\ref{eq:Szfin}), with
\begin{equation}
  g_{{\rm wp},r} \, = \, -\frac{1}{\rho_0}
  \frac{\partial U_s}{\partial r}  \,\, .
  \label{eq:gwpr0}
\end{equation}
Note that $U_s$ is constant in the limit of small-amplitude
acoustic-gravity waves that propagate (i.e., with $\omega > \omega_{a}$)
in a hydrostatic, isothermal, and plane-parallel atmosphere.
In that case, $g_{{\rm wp},r} = 0$.
Dissipation is thus required for a net upward wave-pressure
acceleration, and the wave action conservation equation derived in
{\S}~3.3 gives the required value of $\partial U_{s} / \partial r$.
Equation (\ref{eq:wasimp}) can be used to obtain a good approximation
for the dissipation rate, and
\begin{equation}
  g_{{\rm wp},r} \, = \, \frac{\omega \, T \Delta S}{2\pi c_s}
  \, \approx \, \frac{T \Delta S}{4\pi H}  \,\, .
  \label{eq:gwpr1}
\end{equation}
The use of equation (\ref{eq:wasimp}) implies that only the direct
shock dissipation is being considered as a contributor to a
nonzero wave energy gradient.
It may {\em not} be correct to apply the more complete equation of
wave action conservation (eq.~[\ref{eq:wafull}]) to the expression
for wave pressure, since the additional radial derivatives of
$A$, $V_{\rm gr}$, and $V_{\rm ph}$ are not dissipative in origin.
These other contributors to $\partial U_{s} / \partial r$ are
essentially there to keep the wave energy flux locally conserved,
and they do not represent irreversible energy exchange with the
surrounding medium.

Note that the right-most approximation in equation (\ref{eq:gwpr1})
makes the assumption that $\omega \approx \omega_a$, which
is motivated by the fact that propagating waves are suggested
to arise from resonant excitation (see {\S}~3.2).
The above expression for $g_{{\rm wp},r}$ thus depends
inversely on the density scale height.
In a subsonic atmosphere, one of the effects of wave pressure is
to produce a radial increase in $H$ as the wave amplitude increases.
As $H \rightarrow \infty$, the background medium would become
homogeneous, and it would make sense for
$g_{{\rm wp},r} \rightarrow 0$ because a nonzero wave pressure
gradient relies on there being a background inhomogeneity.
Thus, it is not clear whether the scale height to be used above
should be the value at the piston (which is consistent with the
assumption of constant frequency $\omega$),
or if the locally varying value $H(z)$ would be more consistent
with the known behavior of wave pressure.
Because the proposed Be-star spinup mechanism depends crucially on
the magnitude of ${\bf g}_{\rm wp}$, we aim to be safe and thus
underestimate the local value of $g_{{\rm wp},r}$ by using
the {\em larger} value of the local scale height $H(z)$ in the
denominator of equation (\ref{eq:gwpr1}).

The azimuthal component of ${\bf g}_{\rm wp}$ is not often discussed
in wave-pressure acceleration literature (e.g., as applied to stellar
winds), but it is a key component in studies of the interactions
between waves and mean flows in planetary atmospheres (see {\S}~2).
Equation (\ref{eq:Sxfin}) gives
\begin{displaymath}
  g_{{\rm wp,} \phi} \, = \, 
  -\frac{1}{\rho_0} \left[
  \frac{\partial}{\partial r}
  (\rho_{0} \langle u_{1} w_{1} \rangle) \, + \,
  u_{0} \frac{\partial}{\partial r} (\langle \rho_{1} w_{1} \rangle)
  \,\, +  \right.
\end{displaymath}
\begin{equation}
  \left.
  \langle \rho_{1} w_{1} \rangle \frac{\partial u_0}{\partial r}
  \right] \,\, .
  \label{eq:gwphi}
\end{equation}
The first term, which depends on the phase-averaged product
$\langle u_{1} w_{1} \rangle$, takes the form of a Reynolds stress
that resembles the radial wave pressure (i.e., with
a factor of $u_1$ replacing one of the two factors of $w_1$ implicit
in eq.~[\ref{eq:gwpr0}]).
The second and third terms depend on the phase-averaged product
$\langle \rho_{1} w_{1} \rangle$, and they represent a radial
transport of ``eddy mass flux'' (Lee 2006).
In the theoretical NRP literature, the $\langle u_{1} w_{1} \rangle$
and $\langle \rho_{1} w_{1} \rangle$ effects are associated with
angular momentum transport time scales $\tau_1$ and $\tau_2$
respectively (e.g., Lee \& Saio 1993).

Ando (1983, 1986) found that the $\tau_1$ transport term gives a
net outward transport of angular momentum for prograde modes
($m < 0$, or $k_{x} > 0$) and inward transport for retrograde modes.
This can be understood from the perspective of acoustic-gravity wave
theory by examining the sign dependence of
equation (\ref{eq:u1w1simp}).
The $\tau_1$ transport term can only spin up the outer layers of
a star when $m<0$.
The $\tau_2$ transport term, however, does not depend on the sign of
$m$ at all, and may also be nonzero for purely radial pulsations
($m=0$) in a rotating star.
Lee (2006, 2007) found that for rapidly rotating stars, the $\tau_2$
term is much stronger than the $\tau_{1}$ term near the stellar surface.
This can be verified by using the approximate scaling relations given
in {\S}~3.1 to estimate the relative strengths of the two effects.
The ratio of the first of the two $\tau_2$ terms in equation
(\ref{eq:gwphi}) to the $\tau_1$ term is given roughly by
\begin{equation}
  \frac{\rho_{1} u_{0}}{\rho_{0} u_{1}} \, \approx \,
  \frac{\omega u_0}{c_{s}^{2} k_x} \, \approx \,
  \left( \frac{u_0}{c_s} \right)
  \left( \frac{1}{2 k_{x} H} \right)
  \, \gg \, 1  \,\, .
\end{equation}
The third term in equation (\ref{eq:gwphi}) acts somewhat like an
effective viscosity.
It was derived under the assumption of a plane-parallel geometry
to be proportional to $\partial u_{0} / \partial r$, but in a
spherical geometry it should in fact be proportional to
the full radial gradient of the angular momentum $r u_0$.
This generalization is included in equation (\ref{eq:gwpp1}).

In the models presented below, the amplitude and phase relations
derived in {\S}~3.1 are used to scale the azimuthal wave pressure
with the radial wave pressure.
Thus, the expression used is
\begin{equation}
  g_{{\rm wp,} \phi} \, = \, (\Phi_{1} + \Phi_{2}) g_{{\rm wp,} r}
  - \frac{\langle \rho_{1} w_{1} \rangle}{\rho_0}
  \left( \frac{\partial u_0}{\partial r} + \frac{u_0}{r} \right)
  \label{eq:gwpp1}
\end{equation}
where
\begin{equation}
  \Phi_{1} \, = \,
  \frac{\langle u_{1} w_{1} \rangle}{\langle w_{1}^{2} \rangle}
  \,\, , \,\,\,\,
  \Phi_{2} \, = \,
  \frac{u_{0} \langle \rho_{1} w_{1} \rangle}
  {\rho_{0} \langle w_{1}^{2} \rangle}  \,\, .
  \label{eq:Phis}
\end{equation}
Note that when acoustic-gravity waves are evanescent,
both $u_1$ and $w_1$ are {90\arcdeg} out of phase with one another
(making $\Phi_{1} = 0$) and
$\rho_1$ and $w_1$ are {90\arcdeg} out of phase
(making $\Phi_{2} = 0$).
When applying these transport terms to stellar interiors,
Ando (1983, 1986), Lee \& Saio (1993), and Lee (2006, 2007)
depended on nonadiabatic effects to produce departures from
these {90\arcdeg} phase differences.
In the present case, however, these subtle effects do not need
to be invoked.
{\em Radially propagating} waves in stellar atmospheres naturally
exhibit $\Phi_{1} \neq 0$ and $\Phi_{2} \neq 0$.

\section{Conservation Equations}

In order to evaluate how the wave interactions derived in {\S}~3 
impact the overall atmosphere, both the mean and fluctuating
fluid properties must be computed simultaneously.
The time-averaged density and flow speeds obey conservation equations
for mass, radial momentum, and angular momentum that are given below.
These equations are solved for {\em time-independent} fluid properties
with the implicit assumption that the wave pressure terms---despite
having their origin in temporally fluctuating motions---are essentially
time-steady.
The energy conservation equation is assumed to be satisfied by a
known and constant temperature $T$.

The mass conservation equation for a spherically symmetric
atmosphere can be written in two ways, depending on whether one
tracks the mass flux through a fixed location in the star's inertial
frame, or whether individual fluid parcels are followed in time.
The former (Eulerian) treatment yields the most commonly seen version
of the equation of mass conservation,
\begin{equation}
  \dot{M} \, = \, 4\pi \rho_{0} w_{0} r^{2}
  \label{eq:mcons}
\end{equation}
where $\dot{M}$ is the mass loss rate averaged over the entire star
and $r$ is measured from the center of the star.
Alternately, equation (\ref{eq:massLag}) can be used to determine
the more complete (Lagrangian) mass flux, with
\begin{equation}
  \dot{M} \, = \, 4\pi \left(
  \rho_{0} w_{0} + \langle \rho_{1} w_{1} \rangle
  \right) r^{2}  \,\, .
  \label{eq:mconL}
\end{equation}
The proper physical interpretation of equation (\ref{eq:mconL}),
however, is not clear.
In the photospheres of pulsating stars (including the Sun), the
magnitude of the so-called ``Stokes drift'' term
$\langle \rho_{1} w_{1} \rangle$ may exceed the time-steady mass
flux $\dot{M} / (4\pi r^{2})$ by at least an order of magnitude.
For example, in the unified photospheric, chromospheric, and
coronal models of Cranmer et al.\  (2007), the maximum of the
Stokes drift velocity
$\langle \rho_{1} w_{1} \rangle / \rho_0$ for the Sun occurs in
the upper chromosphere with a magnitude of $\sim$7 km s$^{-1}$
(about half the local sound speed), and at this height the
mean upflow speed of the solar wind (computed effectively by
solving eq.~[\ref{eq:mcons}] for $w_0$) is only 0.4 km s$^{-1}$.
If equation (\ref{eq:mconL}) were the correct time-steady
mass conservation equation, it would imply that the Eulerian radial
velocity $w_0$ would have to be {\em negative} in these
regions---i.e., approximately equal in magnitude to the Stokes
drift velocity, but opposite in sign---to nearly cancel out the
Stokes drift and produce the smaller known mass flux.
This appears to be a kind of unphysical ``fine tuning'' that tends
to appear in situations where some key piece of physics has been
neglected.
On the other hand, equation (\ref{eq:mconL}) does seem to be a
more self-consistent treatment of the second-order wave pressure
terms as derived in the Appendix.

In the models below, we take an agnostic approach to this dilemma
and simply choose the version of the mass flux conservation
equation that produces a {\em smaller} angular momentum transport
in the upper atmosphere of a Be star.
If this choice was the wrong one to make on the basis of physical
realism, then correcting it would only {\em increase} the ability
of NRP/wave coupling to be an effective agent in spinning up
Be-star disks.
The more ``conservative'' choice was found to be
equation (\ref{eq:mcons}) rather than equation (\ref{eq:mconL});
see below.
The former equation is solved with the mass loss rate $\dot{M}$
specified with a known constant value and the density stratification
$\rho_{0}(r)$ constrained by the radial momentum equation.
Equation (\ref{eq:mcons}) is thus solved for the local value of
the Eulerian outflow speed $w_0$.

The radial momentum equation is given in the subsonic limit
(i.e., $w_{0} \ll c_{s}$) as
\begin{equation}
  \frac{c_{s}^2}{\gamma \rho_0} \frac{\partial \rho_0}{\partial r}
  \, = \, -\frac{GM_{\ast}}{r^2} + \frac{u_{0}^2}{r}
  + g_{\rm rad} + g_{{\rm wp},r}
  \label{eq:wcons}
\end{equation}
where the first two terms on the right-hand side (gravity and the
centrifugal force) are given in their full spherical form.
As discussed above, we limit ourselves to isothermal fluctuations
having $\gamma = 1$.
The radiative acceleration $g_{\rm rad}$ is specified using the
numerical model atmosphere results of Lanz \& Hubeny (2007).
The photospheric value of $g_{\rm rad}$ is assumed to apply for
the entire atmospheric model, and thus the additional dependences
on the density and the velocity gradient that apply to stellar winds
(Castor et al.\  1975) are not applied here.

Figure 4 illustrates a parameterized fit to the Lanz \& Hubeny (2007)
radiative acceleration results as a function of stellar gravity
and effective temperature (compare with their Figure 19).
The so-called Eddington-limit acceleration $g_{\rm edd}$ is the
surface gravity at which a star of a given $T_{\rm eff}$ has
sufficient radiative acceleration to completely cancel out its
gravity.
The effective temperature dependence of this quantity has been
fit with a power law, with
\begin{equation}
  g_{\rm edd} \, = \, 142.35  \left(
  \frac{T_{\rm eff}}{2 \times 10^{4} \,\, \mbox{K}} \right)^{3.8575}
  \,\,\, \mbox{cm} \,\, \mbox{s}^{-2}  \,\, .
  \label{eq:gedd}
\end{equation}
The resulting radiative acceleration, expressed as a dimensionless
ratio to the surface gravity ($g > 0$), was then fit with the
following polynomial relation,
\begin{equation}
  \ln \left( \frac{g_{\rm rad}}{g} \right) \, = \,
  -0.25943 x - 0.037106 x^{2}
  \label{eq:grad}
\end{equation}
where $x \equiv \ln ( g / g_{\rm edd} )$.
This expression is valid only for $g > g_{\rm edd}$ (i.e., below
the Eddington limit).
Figure 4 also shows how $\log g$ depends on $T_{\rm eff}$ for
standard definitions of stars having luminosity classes III, IV,
and V, using the spectral type calibrations of
de Jager \& Nieuwenhuijzen (1987) and Cranmer (2005).
In the Be-star models below, we use the centrifugally-modified
gravity (i.e., the first two terms on the right-hand side of
eq.~[\ref{eq:wcons}]) for $g$ in the parameterization for
$g_{\rm rad}$.

\begin{figure}
\epsscale{1.12}
\plotone{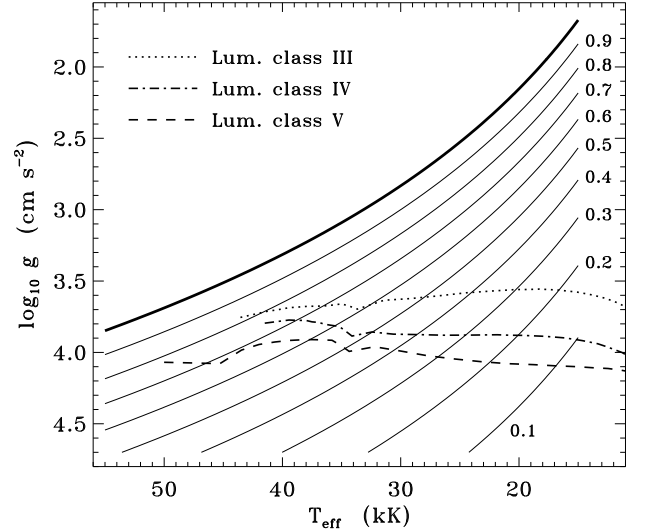}
\caption{Contours of constant values for the ratio of radiative to
gravitational acceleration, as a function of stellar surface gravity
$g$ and effective temperature $T_{\rm eff}$, computed from fits
to the numerical models of Lanz \& Hubeny (2007).  Values of the
ratio $g_{\rm rad}/g$ are listed next to each curve, with the
Eddington limit ($g_{\rm rad}/g = 1$) shown as a thicker curve.
Also shown are locations of luminosity classes III ({\em dotted line}),
IV ({\em dot-dashed line}), and V ({\em dashed line}) stars in
$g, T_{\rm eff}$ space.}
\end{figure}

In practice, it is simpler to solve equation (\ref{eq:wcons}) for the
local scale height $H$ than to integrate it directly to obtain the
density $\rho_{0}$.
Using equation (\ref{eq:gwpr1}) for the wave-pressure acceleration,
the scale height is given by
\begin{equation}
  H \, = \, \frac{c_{s}^{2} + (T \Delta S / 4\pi)}{g_{\rm eff}}
  \label{eq:Hcons}
\end{equation}
where the effective gravity
\begin{equation}
  g_{\rm eff} \, = \,
  \frac{GM_{\ast}}{r^2} - \frac{u_{0}^2}{r} - g_{\rm rad}
\end{equation}
is always positive.
Note that the scale height can become {\em large} in two interesting
limiting cases:
(1) when the shock dissipation $T \Delta S$ is substantial,
and (2) when the atmosphere spins up enough so that the effective
gravity $g_{\rm eff}$ is small. 

The azimuthal component of the momentum conservation equation, which
determines the bulk rotation speed $u_0$, is
\begin{equation}
  w_{0} \left( \frac{\partial u_0}{\partial r} +
  \frac{u_0}{r} \right) \, = \, g_{{\rm wp},\phi}  \,\, .
  \label{eq:ucons}
\end{equation}
No horizontal radiative acceleration is assumed here, though it may
be present in the supersonic regions of rapidly rotating winds or
disks (e.g., Grinin 1978; Owocki et al.\  1996; Gayley et al.\  2001;
Kub\'{a}t 2007).
Equation (\ref{eq:gwpp1}) is used for the azimuthal wave-pressure
acceleration $g_{{\rm wp},\phi}$, and thus the conservation equation
can be rearranged as
\begin{equation}
  \left( \rho_{0} w_{0} + \langle \rho_{1} w_{1} \rangle \right)
  \left( \frac{\partial u_0}{\partial r} +
  \frac{u_0}{r} \right) \, = \, (\Phi_{1} + \Phi_{2})
  \rho_{0} g_{{\rm wp},r} \,\, .
  \label{eq:u2solv}
\end{equation}
It is clear that {\em without} any horizontal wave pressure,
the equilibrium solution for the rotation speed would be
$u_{0} \propto r^{-1}$; i.e., angular momentum conservation and
no spinup to a Keplerian disk.
The mass flux given in the first set of parentheses above is
identical to the Lagrangian terms from equation (\ref{eq:massLag}).
When deciding which form of the mass flux conservation should be
applied in these models (i.e., eq.~[\ref{eq:mcons}] or
[\ref{eq:mconL}]) it was concluded that the safer choice would be
to use the version that gives the {\em larger} mass flux.
In that case, the right-hand side of equation (\ref{eq:u2solv})
would be divided by a larger quantity, and the radial gradient
of the angular momentum would be increased by a smaller (i.e.,
more conservatively underestimated) amount.
As mentioned above, this choice was found to be equation
(\ref{eq:mcons}), since in that case both $w_0$ and the Stokes
drift velocity $\langle \rho_{1} w_{1} \rangle / \rho_0$ are
always positive.

Equations (\ref{eq:wafull}), (\ref{eq:mcons}), (\ref{eq:Hcons}),
and (\ref{eq:u2solv}) were solved simultaneously
by a numerical code that integrates upwards from the photospheric
lower boundary.
The numerical quadrature was done with straightforward explicit
Euler integration steps.
The density $\rho_0$ was advanced along the radial grid by using
the local value of the scale height $H$.
Most runs of the code utilized 4000 radial grid zones that covered
80 photospheric scale heights, and tests with double the number of
grid zones verified that the resolution was adequate.

Because the radial momentum conservation (eq.~[\ref{eq:wcons}])
does not account self-consistently for the
transonic and supersonic parts of the stellar wind, additional
checks are made in the code to ensure that the subsonic assumption
is not violated.
If the integrated value of $w_0$ exceeds a specified terminal
speed $w_{\infty}$, then the wind speed is not allowed to increase
any further and the density scale height is set to an asymptotic
value $H_{\infty} = r/2$.
These solutions are judged to be ``winds'' and not ``disks.''
Furthermore, the wave amplitude $w_1$ is monitored so that it does
not exceed $V_{\rm max}$ (which scales with $\rho^{-1/2}$ and is
normalized to the photospheric boundary condition for $w_1$).
This ensures that the modeled resonant excitation does not extract
more energy from the evanescent waves than is available.

\section{Model Results}

\subsection{Example B2 V Star at 70\% Critical Rotation}

Figure 5 shows the result of integrating the conservation equations
derived above for an example B2~V star rotating well below its
critical rotation speed.
For this star, $M_{\ast} = 7.43 \, M_{\odot}$ and
$\log L_{\ast}/L_{\odot} = 3.47$, and the polar radius was assumed
to be $R_{p} = 4.12 \, R_{\odot}$ (see Cranmer 2005).
The equatorial rotation speed $V_{\rm rot}$ was chosen to be 70\%
of the critical speed, with $V_{\rm crit} = 479$ km s$^{-1}$ for
this star.
Assuming rigid rotation and Roche equipotentials, the equatorial
radius is $R_{e} = 4.92 \, R_{\odot}$.
Standard von Zeipel (1924) gravity darkening gives the equatorial
$T_{\rm eff}$ to be 17920 K, which is lower than both the
corresponding polar value of 22170 K and the nonrotating value of
20900 K.
The equatorial scale height at the photosphere is
$H_{0} = 0.0017 \, R_{e}$, and the maximum height shown in
Figure 5 is $80 H_{0}$, or $\sim$0.14 $R_{e}$.

\begin{figure}
\epsscale{1.08}
\plotone{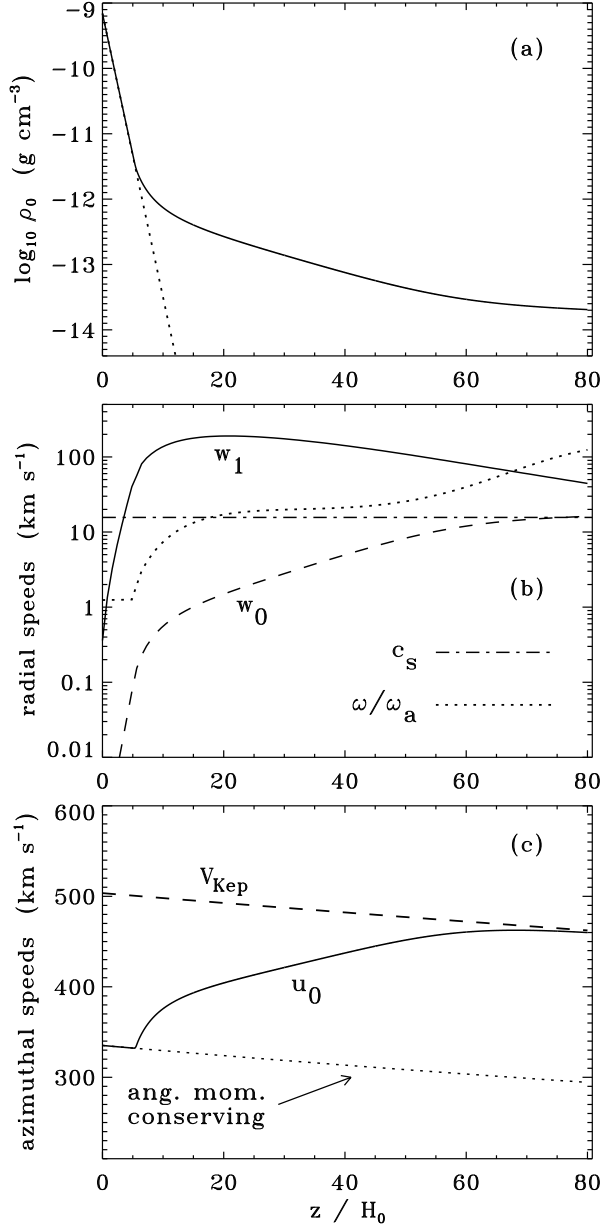}
\caption{Radial dependence of atmospheric parameters for
example B2~V star:
({\em{a}}) mean density $\rho_0$ modeled with wave pressure
({\em solid line}) and without ({\em dotted line}).
({\em{b}}) bulk wind velocity $w_0$ ({\em dashed line}),
radial wave velocity amplitude $w_1$ ({\em solid line}), sound
speed $c_s$ ({\em dot-dashed line}), and ratio of resonant wave
frequency to local acoustic cutoff frequency ({\em dotted line}).
({\em{c}}) bulk azimuthal speed $u_0$ modeled with wave pressure
({\em solid line}) and without ({\em dotted line}), and the
local Keplerian rotation speed ({\em dashed line}).}
\end{figure}

The photospheric boundary condition on density $\rho_0$ is
computed from a tabulated grid of Rosseland mean opacities
$\kappa_{\rm R}$ (Kurucz 1992) and the
condition that $\tau \approx \kappa_{\rm R} \rho_{0} H_{0} = 1$.
The value for the example star, at the equator, was computed to
be approximately $7 \times 10^{-10}$ g cm$^{-3}$.
This model uses a representative B2~V star mass-loss rate
$\dot{M} = 10^{-9} \, M_{\odot}$ yr$^{-1}$, which assumes that the
polar wind's mass-loss rate (estimated here using the fitting
relations of Vink et al.\  2000) remains comparable to the equator's
mass loss rate (see {\S}~5.1 of Bjorkman 2000).
Also, we assume a stellar wind terminal speed $w_{\infty} = 1500$
km s$^{-1}$.
Note, however, that the adopted values of $\dot{M}$ and $w_{\infty}$
tend not to have any real impact the quantitative properties of
``successful'' disk models---i.e., models that do {\em not} exceed
the supersonic wind condition ($w_{0} = w_{\infty}$)---because
the $w_0$ term in equation (\ref{eq:u2solv}) is dominated in those
cases by the viscosity-like (Stokes drift) term.

The example stellar model shown in Figure 5 was assumed to have a
driving NRP period of 10 hours, which is equivalent to an evanescent
frequency ratio $\omega / \omega_{a} = 0.13$ in the photosphere
(see Fig.~1).
This frequency enters into the model calculations only in the
factor of $(\omega_{a}^{2} - \omega^{2})$ in the denominator of
equation (\ref{eq:Vfree}), so the model is relatively insensitive
to the exact value of the driving frequency.
Elsewhere in the model, the wave frequency $\omega$ is taken to be
that of the resonantly excited waves at the subphotospheric piston.
The acoustic cutoff period at the piston is 1.07 hours, and this
is slightly shorter than the local acoustic cutoff period in the
photosphere (the latter using $g_{\rm rad}$ to lower the effective
gravity), which is 1.32 hours.
The adopted photospheric NRP amplitude is 20 km s$^{-1}$ (i.e.,
1.28 times the sound speed), and the reflection height
$\Lambda = 10 H_0$.
This being a model for a Be star, the azimuthal mode is assumed
to be retrograde, with $\ell = m = 2$ (Rivinius et al.\  2003),
and thus $k_{x} H_{0} = -0.0035$.

Figure 5{\em{a}} shows how the inclusion of wave pressure increases
the scale height and gives rise to a significantly shallower density
gradient.
The velocity amplitude $w_1$ of the resonantly excited wave is shown
in Figure 5{\em{b}}, and the initial quasi-exponential rise
(eq.~[\ref{eq:Vfree}]) is halted by shock dissipation.
The radial dependence of the time-steady wind speed $w_0$ is
essentially the reciprocal of the density.
Figure 5{\em{b}} shows that for this model $w_0$ just begins to
exceed the sound speed $c_s$ at the top of the spatial grid, but it
is also evident that without wave pressure this would have
occurred at a much lower height ($z/H_{0} \sim 10$).
The angular momentum deposition from $g_{{\rm wp},\phi}$ is
evident in Figure 5{\em{c}}, which shows the abrupt appearance of
nonzero wave pressure beginning at the height where shocks first
occur.
The angular momentum transport saturates as $u_{0}$ approaches
$V_{\rm Kep}$ because in that limit $g_{\rm eff} \rightarrow 0$
and $H \rightarrow \infty$.

The critical rotation speed $V_{\rm crit}$ is not the same quantity
as the Keplerian speed $V_{\rm Kep}$.
The latter is the rotation speed needed for a given star to be
able to efficiently spin material up into orbit and form a disk.
Formally, both $V_{\rm crit}$ and $V_{\rm Kep}$ are defined as the
rotation speeds required to balance gravity, but $V_{\rm crit}$
presupposes that the whole star expands out to a critical equatorial
radius of $R_{e} \approx 1.5 \, R_{p}$.
In the numerical models presented here, $R_e$ is held fixed at the
value determined by the bulk stellar rotation (at and below the
photosphere) and $V_{\rm Kep} \equiv (g r)^{1/2}$ is the
speed required to balance gravity at that radius.
For this star, the condition $V_{\rm rot}/V_{\rm crit}=0.70$
corresponds to a ratio $V_{\rm rot}/V_{\rm Kep}=0.665$.
At the photosphere, $u_{0} = 335$ km s$^{-1}$ and
$V_{\rm Kep} = 504$ km s$^{-1}$, and thus the atmosphere must spin up
by at least $10 c_s$ in order to form a Keplerian disk.
Figure 5 shows that this model contains sufficient angular momentum
transport to accomplish this spinup {\em below} the sonic point of
the radial outflow.

\subsection{Varying the Stellar Parameters}

The model described above serves as an illustration of how resonant
excitation, shock dissipation, and wave pressure may act together to
spin up the outer layers of a rotating star.
This outcome, though, depends on several key parameters
that we vary below to explore whether this process can truly be an
explanation for observed Be-star disks.

Figure 6 shows contours of various scalar properties taken from
a two-dimensional grid of models that vary both the rotation speed
$V_{\rm rot}$ (vertical axis) and the vertical NRP velocity
amplitude $V_{\rm puls}$ (horizontal axis) at the photosphere of
the B2~V star.
The grid contained 150 values of $V_{\rm rot}/V_{\rm crit}$
distributed linearly between 0.01 and 0.995, and 150 values of
$V_{\rm puls}$ distributed logarithmically between 0.1 and 100
km s$^{-1}$.
All other parameters were held fixed at the values given in {\S}~5.1.
The quantities shown in Figures 6{\em{a--c}} are measured at the
top of the spatial grid ($z = 80 H_{0}$).
Figure 6{\em{a}} shows whether the wave pressure was able to
increase the bulk azimuthal speed $u_0$ to the local Keplerian
speed $V_{\rm Kep}$ by the top of the grid.
It is clear that a star requires both a relatively high photospheric
rotation speed (so that $u_0$ does not have so far to go) and a
substantial pulsation amplitude to deposit sufficient angular momentum.
However, this appears to be able to occur for some cases when
$V_{\rm rot}$ is as small as about $0.6 V_{\rm crit}$ (for larger
NRP amplitudes), or when $V_{\rm puls}$ is as small as
1 or 2 km s$^{-1}$ (for larger rotation rates).
Figure 6{\em{b}} shows the radial wind speed $w_0$ in units of
the adopted terminal speed $w_{\infty}$.
There is a rough correlation between efficient spinup to Keplerian
rotation and a {\em low} wind speed at the top of the grid.

\begin{figure}
\epsscale{1.15}
\plotone{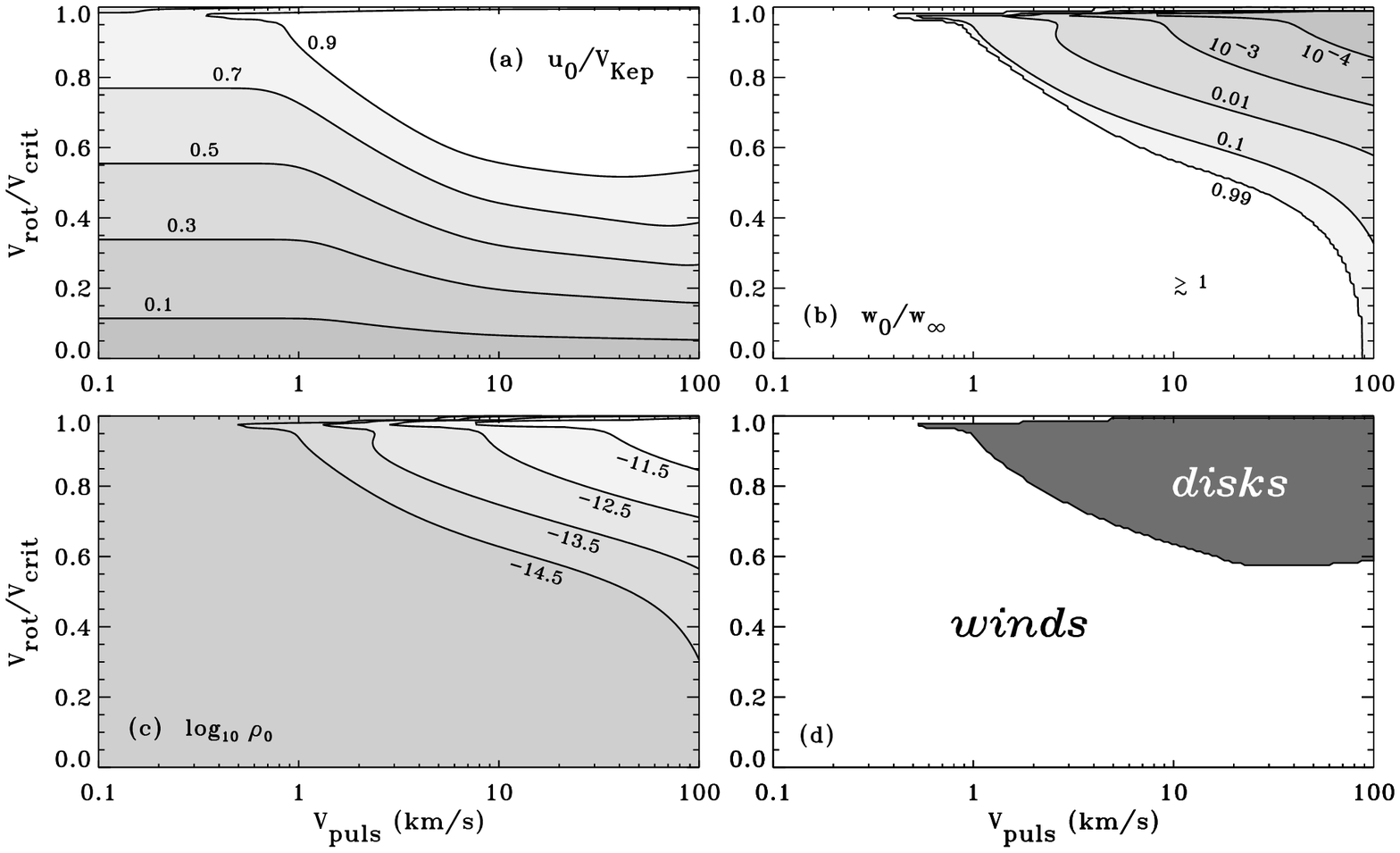}
\caption{Properties at the largest height ($z = 80 H_0$) for
a grid of B2~V star models that vary $V_{\rm rot}$ and
$V_{\rm puls}$ independently of one another:
({\em{a}}) ratio of rotation speed $u_0$ to Keplerian orbital
speed $V_{\rm Kep}$,
({\em{b}}) ratio of radial wind speed $w_0$ to assumed terminal speed
$w_{\infty}$, and
({\em{c}}) mean mass density $\rho_0$.
Labels for constant contour values are given next to each curve.
Panel ({\em{d}}) shows the heuristic dividing line between winds
and disks (see text).
{\bf See the last page of this paper for the full two-column
version of this figure.}}
\end{figure}

Figure 6{\em{c}} shows contours of the mean density $\rho_0$ at
the top of the radial grid, which we take as a proxy for the
density at the ``inner edge'' of the decretion disk.
(For the models exhibiting Keplerian rotation, the density gradient
is shallow and the plotted values of $\rho_0$ are insensitive
to the exact height assumed for the upper edge of the grid.)
The densities in Figure 6{\em{c}} range between a minimum of 
$3 \times 10^{-16}$ g cm$^{-3}$ (lower left) and a maximum of
$8 \times 10^{-11}$ g cm$^{-3}$ (upper right), with the values
exceeding $10^{-14}$ g cm$^{-3}$ tending to occur in the regime of
parameter space filled by Keplerian disks.
This range is consistent with observed inner disk densities.
Traditional measurements from infrared excess (Waters et al.\  1987)
and visible linear polarization (McDavid 2001) give values
between about $10^{-12}$ and $1.5 \times 10^{-11}$ g cm$^{-3}$.
More recent determinations that make use of H$\alpha$ profiles
and interferometry (e.g., Gies et al.\  2007; Jones et al.\  2008)
exhibit values that vary a bit more widely; essentially from
$\sim$10$^{-13}$ to $10^{-9}$ g cm$^{-3}$.
The latter end of this range, however, overlaps with estimated
photospheric densities, and thus these values may be inconsistent
with observational evidence that the photospheric layers rotate
more slowly than $V_{\rm Kep}$.

Figure 6{\em{d}} illustrates the ``bifurcation'' of the models
between winds and disks.
For the purposes of this dividing line, a model is considered to
have produced a Keplerian disk only if both
$u_{0}/V_{\rm Kep} \geq 0.95$ and $w_{0}/w_{\infty} < 0.1$.
The latter condition ensures that an equatorial wind has not
yet accelerated significantly by the top of the grid, which
would invalidate the neglect of supersonic terms in
equation (\ref{eq:wcons}).

Figure 7 shows a similar set of contours as Figure 6{\em{a}}, but
the horizontal axis varies the reflection height $\Lambda$
as a free parameter (see {\S}~3.2).
The two-dimensional grid contained 120 values of
$V_{\rm rot}/V_{\rm crit}$ distributed linearly between 0.01 and
0.995, and 200 values of $\Lambda/H_0$ distributed logarithmically
between 1 and 500.
This figure also includes the bifurcation curve as defined for
Figure 6{\em{d}}.
The photospheric NRP velocity amplitude was held fixed at
20 km s$^{-1}$.
The combined effects of resonant excitation, shock dissipation, and
wave pressure appear to produce roughly the same amount of angular
momentum transport for $1 \lesssim \Lambda/H_{0} \lesssim 20$.
The ``kink'' in the contours at $\Lambda/H_{0} \approx 3.5$
corresponds to the reflection height at which the efficiency of
resonant excitation saturates to unity somewhere above the
photosphere (see Fig.~3).
The largest values of the reflection height
($\Lambda/H_{0} \gtrsim 60$) do not give enough resonant
excitation to produce substantial angular momentum transport
in the upper atmosphere, so the contours approach the same kind
of limiting shapes as are seen in Figure 6 for the limit
$V_{\rm puls} \rightarrow 0$.

\begin{figure}
\epsscale{1.12}
\plotone{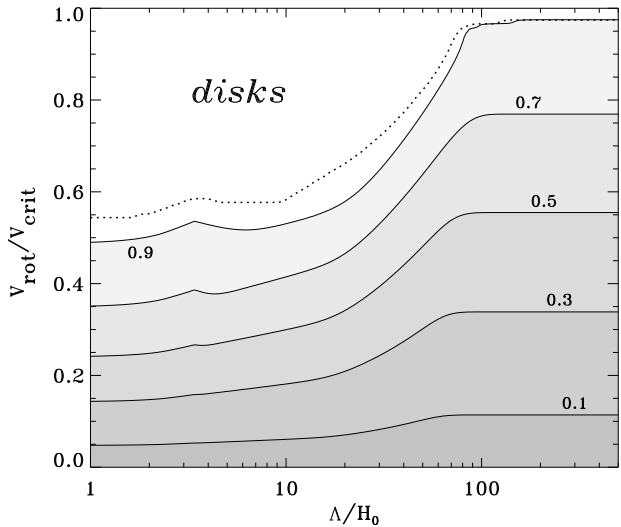}
\caption{Contours of $u_{0}/V_{\rm Kep}$ plotted similarly to
Figure 6{\em{a}}, but for a grid of B2~V star models that vary
$V_{\rm rot}$ and the reflection height $\Lambda$ independently
of one another.  The photospheric pulsation amplitude is held
fixed at 20 km s$^{-1}$.  The disk/wind boundary from
Figure 6{\em{d}} is also shown ({\em dotted line}).}
\end{figure}

The final grid of models was produced by varying the rotation
rate and the {\em spectral type} of the star, while keeping both
$V_{\rm puls}$ and $\Lambda$ fixed at their standard values
of 20 km s$^{-1}$ and $10 H_0$, respectively.
The grid contained 200 values of $V_{\rm rot}/V_{\rm crit}$
distributed linearly between 0.01 and 0.995, and 55 values of
the spectral type in half-subtype increments between O3 and F0.
Figure 8 shows contours of both $u_{0} / V_{\rm Kep}$
(Fig.~8{\em{a}}) and $w_{0} / w_{\infty}$ (Fig.~8{\em{b}})
along with the wind/disk bifurcation curve that is constrained
by the combination of these two quantities.
As before, the stellar parameters for each spectral type were
taken from the $L_{\ast}$ and $T_{\rm eff}$ calibration of
de Jager \& Nieuwenhuijzen (1987), and $M_{\ast}$ was
determined from the evolutionary tracks of Claret (2004).
Fits for the mass-luminosity relations for luminosity classes III,
IV, and V were given by Cranmer (2005).
The models shown in Figure 8 follow the main sequence (i.e.,
luminosity class V).

\begin{figure}
\epsscale{1.12}
\plotone{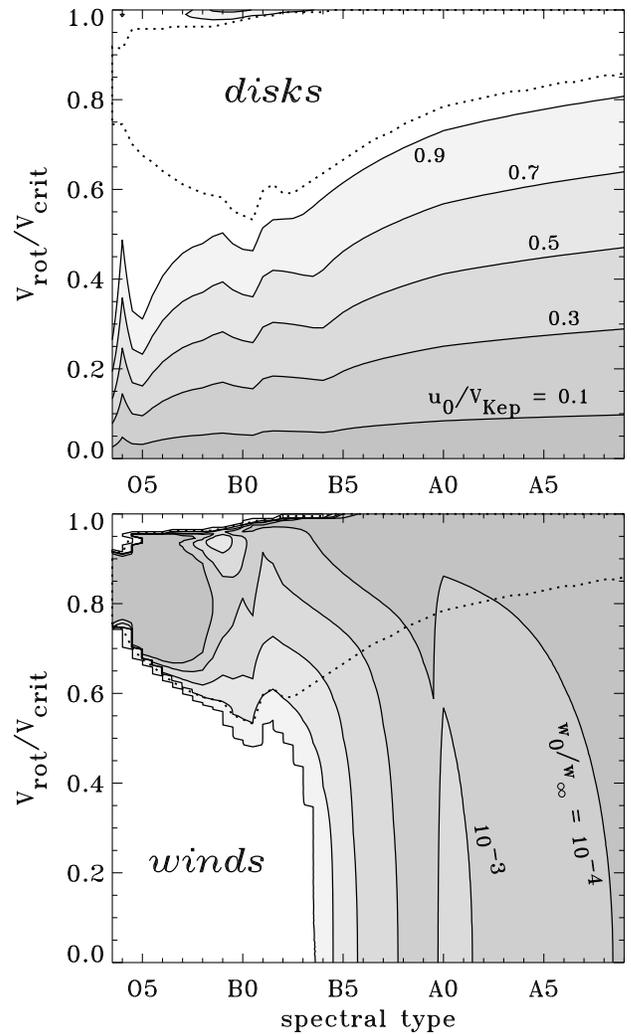}
\caption{Contours of $u_{0}/V_{\rm Kep}$ ({\em{top}}) and
$w_{0}/w_{\infty}$ ({\em{bottom}}) plotted similarly to
Figures 6{\em{a--b}}, but for a grid of main sequence models that
vary $V_{\rm rot}$ and the spectral type independently of one
another.  The disk/wind boundary from Figures 6{\em{d}} and 7 is
also shown in each panel ({\em dotted lines}).}
\end{figure}

The position of the dividing line between winds and disks depends
(somewhat) on assumptions made about the equatorial mass loss rate
and terminal speed.
The fitting relations of Vink et al.\  (2000) were used to
estimate $\dot{M}$ as a function of the stellar parameters.
The terminal speed was assumed to scale as
$w_{\infty} = 1.8 V_{\rm esc}$, with the escape speed $V_{\rm esc}$
calculated at the pole (i.e., to estimate the polar wind's
terminal speed).
The choice of the numerical constant 1.8 was motivated to
reproduce the value of $w_{\infty} = 1500$ km s$^{-1}$ used in
the standard B2~V model discussed above.
Although this constant has been seen to be as large as $\sim$3 for
earlier O stars (Prinja et al.\  1990), the shapes of the contours
in Figure 8 were not seen to change when this constant was varied
between 1.8 and 3.

Figure 8 pinpoints early B-type stars as those most likely to form
Keplerian decretion disks with largest range of possible photospheric
rotation rates.
The O-type stars have stronger winds that require a more
self-consistent (subsonic to supersonic) treatment of the
radial momentum equation.
Late B-type and all A-type stars appear to require only slightly larger
rotation rates than early B stars to produce disks, but one should
recall that the NRP amplitude was held fixed at 20 km s$^{-1}$ across
these spectral types.
In fact, most ``normal'' A stars (along and above the main sequence)
are not as pulsationally unstable as the $\beta$~Cep and SPB stars,
so their NRP amplitudes are likely to be significantly smaller than
assumed here.

\section{Discussion}

\subsection{Time Steady Disk Formation}

How believable are the models presented above?
This is a nontrivial question to ask when attempting to solve a
problem that has plagued a discipline for three quarters of a century.
The chain of events described in this paper remains somewhat
speculative mainly because of the need to assume a finite
reflection time (or reflection height $\Lambda$) that provides
a time-steady yield of resonantly excited waves.
If the resonant waves are much weaker than those predicted with
$\Lambda \lesssim 20 H_{0}$ (see Fig.~7), then this pulsational
mechanism may not work.
Additionally, if the frequency of resonant oscillations varies
with height to stay equal to the {\em local} acoustic cutoff
frequency (rather than remain at the piston's cutoff frequency),
then the waves may not propagate upward.
In that case, the phase shift factors $\Phi_1$ and $\Phi_2$
(eq.~[\ref{eq:Phis}]) could be much smaller than if the waves were
propagating, and the ability to transport angular momentum would
be reduced.

It is important to note, though, that there were several places in
the above analysis where the wave coupling effects were decidedly
{\em underestimated.}\footnote{%
In part, this was done to counteract some of the ``irrational
exuberance'' that an advocate for a new theoretical model is likely
to exhibit!}
First, equation (\ref{eq:Vfree}) neglects all higher-order transient
terms (i.e., $t^{-5/2}$, $t^{-7/2}$, and so on) in the series
expansion for the resonantly excited wave power.
Second, equation (\ref{eq:gwpr1}) uses the local scale height
$H(z)$ in the denominator of the wave pressure, rather than the
smaller value at the piston that would have been more consistent
with the use of a constant frequency $\omega$.
Third, the Eulerian mass conservation equation (eq.~[\ref{eq:mcons}])
was solved for $w_0$ instead of the Lagrangian version, resulting
in a potentially weaker impact on the angular momentum transport.
Fourth, this analysis totally neglected the fact that even purely
evanescent waves can propagate energy in the horizontal direction
(Walterscheid \& Hecht 2003), and thus they may be able to exert a
substantial wave pressure force in the azimuthal direction---even
if they do not convert any of their energy into the resonant mode.

This paper made the implicit assumption that many early-type
Be stars are rotating significantly below the critical limit, as
suggested by, e.g., Slettebak (1982), Mennickent et al.\  (1994),
Porter (1996), and Cranmer (2005).
If instead {\em all} Be stars are rotating close to critically
(i.e., $V_{\rm rot}/V_{\rm crit} \gtrsim 0.95$) as suggested by
Maeder \& Meynet (2000) and Townsend et al.\  (2004),
the formation of disks could be accomplished with much more
inefficient---and less speculative---physical mechanisms
(see also Owocki 2005).
In any case, for Be stars undergoing NRP with amplitudes of order
the sound speed in their photospheres, many of the processes
described in this paper should be occurring nevertheless.

\subsection{Long-Term Be/Bn Phase Variability}

The most prominent feature of Be stars that has not yet been
addressed here is their substantial variability.
Most noticeably, Be stars are known to exhibit phase changes
between normal B (i.e., ``Bn''), standard Be, and Be-shell
properties (e.g., Slettebak 1988; Kogure 1990).
These changes are typically interpreted as variations in the
density and geometry of the circumstellar disk, though long-term
changes in the properties of the photosphere have also been
inferred (Doazan et al.\  1986).

Although it was suggested earlier ({\S}~1) that a collection of
other proposed mechanisms for producing Be-star disks may be
responsible for some of the observed variability, it is worthwhile
to also examine whether NRP/wave coupling would also contribute.
In the models presented in this paper, the support of Be-star
disks is provided ultimately by the kinetic and thermal
energy in NRP motions.
Thus, variations in the overall properties of the disk could be
caused by variations of the stellar pulsations on the same kinds
of time scales (i.e., long-term changes in the NRP ``piston'').
In fact, rapid stellar rotation has been shown to be connected
with several kinds of pulsational variability and stochasticity:

\begin{enumerate}
\item
The rotational splitting of NRP modes gives rise to closely separated
periods that may be responsible for {\em beating} and amplitude
modulations on time scales of weeks to months.
For example, Rivinius et al.\  (1998) found a set of periods for
$\mu$~Cen that are separated by fractional frequency differences of
order 0.01--0.02, such that the beating maxima and minima can occur
with periods up to several years.
There is also long-term {\em drift} in the precise values of many
observed frequencies that could indicate beating or other kinds
of multi-mode interactions (e.g., \v{S}tefl \& Balona 1996;
Sareyan et al.\  1998).
It has also been suggested (Henrichs 1984) that rotational
splitting---especially for rapid rotation---may create additional
``channels'' of energy transfer between NRP and the mean state, and
may even allow more pulsational energy to escape from the star.
\item
As NRP modes grow to large amplitudes (i.e., $V_{\rm puls} \gtrsim
c_{s}$) a host of nonlinear wave-wave interactions become possible
that are not present in the linear, small-amplitude limit.
There is observational evidence for a given NRP mode to grow slowly
in amplitude, seemingly at the expense of another mode (Smith 1986).
As a pulsation mode becomes nonlinear, it may even lose sinusoidal
coherence by ``interfering with itself'' as it circumnavigates the
star in longitude (e.g., Smith et al.\  1987).
If this occurs it could lead to shock steepening and
dissipation in $\phi$, instead of in $r$ as assumed in {\S}~3.3.
The passage of an azimuthally propagating shock past a given point
on the star could also act as a nonlinear trigger to restart the
wave reflection ``clock'' for resonant excitation.
Osaki (1999) suggested that a combination of these effects---along
with feedback from the dense circumstellar envelope---would
also lead to the eventual quenching of angular momentum transport,
and thus to the dissipation of the disk.
Such a repeated relaxation-oscillation cycle could be responsible
for the phase variations between no disks (Bn), standard disks
(Be), and optically thick disks (Be-shell).
\item
The existence of {\em subsurface convection} could both impact
the long-term growth or decay of pulsation modes and act as a
seed for circumstellar variability (e.g., Cox 1980;
Balmforth 1992; Gabriel 1998; Cantiello et al.\  2009).
Traditionally, B-type stars are not believed to have subsurface
convection zones.
There is some indirect evidence, however, that strong turbulent
motions may be present near the surfaces of rapidly rotating
early-type stars (Smith 1970; Kodaira 1980; Dolginov \& Urpin 1983).
Some recent models of oblate rapid rotators do show the emergence
of a near-surface convection cell at the equator
(Espinosa Lara \& Rieutord 2007; Maeder et al.\  2008).
Even nonrotating hot stars---sufficiently above the main
sequence---may exhibit enhanced subsurface convection due to
newly discovered opacity peaks (Cantiello et al.\  2009).
\end{enumerate}

Although strictly unrelated to the stellar pulsations, another major
source of long-term Be-star variability appears to be the naturally
slow {\em precession} of non-axisymmetric density fluctuations
through the dense Keplerian disk (e.g., Papaloizou et al.\  1992;
Savonije \& Heemskerk 1993; Fi\v{r}t \& Harmanec 2006).
Gas parcels rotating under the influence of a non-point-mass
gravitational potential undergo elliptical orbits with slow
(decade timescale) pattern speeds.
The modes that grow the fastest seem to be $m=1$, or ``one-armed,''
instabilities that appear to match the observed properties of the
evolving violet-to-red ($V$/$R$) ratio of double-peaked Balmer
emission line profiles.
The ability to produce accurate and predictive models of these
variations depends on understanding the origin and supply of
angular momentum at the inner boundaries of the disks.

Finally, magnetic fields should not be neglected as an additional
source of variability.
Recent observational studies are beginning to reveal a distinct
subclass of magnetic Be stars, which includes $\gamma$~Cas and at
least a half-dozen others around spectral types B0e--B1e
(Smith et al.\  2004; Smith \& Balona 2006; Rakowski et al.\  2006;
Lopes de Oliveira et al.\  2007).
Magnetic interactions between the star and the disk could be
the source of dynamo action; i.e., another potential source of
year- to decade-long variations in the circumstellar emission.
Note that these processes may be acting whether or not the stellar
wind is channeled along magnetic flux tubes to feed gas into the
disk (e.g., Cassinelli et al.\  2002; Brown et al.\  2004, 2008;
Ud-Doula et al.\  2008).

\clearpage
\section{Conclusions and Future Prospects}

The primary aim of this paper has been to explore and test a set
of physical processes that may be responsible for the production of
Keplerian decretion disks around classical Be stars.
Although nonradial pulsations correspond to low-frequency
evanescent waves in the photospheres of hot stars, they have
been suggested to give rise to higher frequency resonant
oscillations at the acoustic cutoff frequency.
If these resonant modes grow in amplitude with increasing height,
begin to propagate upwards, and steepen into shocks, the resulting
dissipation would create substantial wave pressure that both
increases the atmospheric scale height and transports angular
momentum upwards.
Using a reasonable assumption for the efficiency of resonant 
wave generation, this chain of events was found to be able to
create the inner boundary conditions required for a dense
Keplerian disk, even when the underlying photosphere is rotating
as slowly as $\sim$60\% of its critical rotation speed.

It is important to note that in the models presented above, the
``spinup'' of the atmosphere occurs high enough {\em above} the
photosphere so that standard measurements of the stellar rotation
speed (i.e., $v \sin i$ from absorption line widths) would not be
sensitive to these motions.
The density at which the spinup is complete occurs several orders
of magnitude below the photospheric ($\tau \approx 1$) density,
and these values are consistent with inner-edge disk densities as
measured by infrared excess and linear polarization techniques
(see {\S}~5.2).
One way of potentially verifying that spinup is occurring
in this thin layer would be to improve on measurements of spectral
lines that are formed at a range of heights above the photosphere
(see, e.g., Chen et al.\  1989).
In addition, the proposed scenario for resonant wave excitation 
implies that the dominant frequency of variability should shift
from low (evanescent piston) to high (acoustic cutoff) values
as the height increases.

In order to further test and verify that NRP-induced wave pressure
is responsible for spinning up Be-star disk material, the physics
of the resonant excitation process needs to be simulated more
accurately.
It may be possible to begin with one-dimensional simulations
(similar to those of Fleck \& Schmitz 1991) to explore both the
time-steady yield of energy conversion from evanescent to
resonant waves and the gradual increase in scale height due to
radial wave pressure.
Also, an analytic non-WKB treatment of partial wave reflection
may shed light on how much intrinsic ``leakage'' the evanescent
waves may undergo, even in the absence of resonant conversion
(see Heinemann \& Olbert 1980; Cranmer \& van Ballegooijen 2005;
Cranmer 2008).

Testing the {\em full} set of processes (e.g., resonant excitation,
wave pressure, and angular momentum transfer) appears to require at
least two-dimensional simulations of the fluid properties in the
equatorial plane of a rotating hot star.
These may take the form of next-generation extensions of studies
by Kroll \& Hanuschik (1997) and Owocki \& Cranmer (2002),
who showed that a large-amplitude NRP-like lower boundary could
give rise to intermittent angular momentum transfer.
Simulations may also benefit from an improved treatment of
viscosity, which would be needed to extend the models
upward to larger distances in the disk midplane (see, e.g.,
Lee et al.\  1991; Okazaki 2001).
It is likely that once simulations reveal the key physical
processes, semi-analytic models like those presented in this
paper can be improved to be more robust, predictively accurate,
and less dependent on free parameters.

For the stellar models in which a supersonic wind begins to
accelerate, a more physically realistic treatment of the radiative
acceleration is needed.
For example, the artificial imposition of a specific value for
$\dot{M}$ should be replaced with a self-consistent calculation
of the disk's mass loss rate, which may be completely uncoupled
from that of the polar wind.
The proper modeling of B-type stellar outflows may also require
an explicit treatment of collisional coupling between the
various atomic and ionic species (Babel 1995;
Krti\v{c}ka \& Kub\'{a}t 2001; Owocki \& Puls 2002;
Votruba et al.\  2007)
and a consistent solution for the internal energy equation,
including both shock heating (e.g., Struck et al.\  2004)
and radiative scattering effects (Gayley \& Owocki 1994).

Lastly, it should be emphasized that future work must involve
not only increased physical realism for the models, but also
quantitative comparisons with observations.
It would be beneficial to apply the methodology outlined in
this paper to a set of well-observed stars, rather than to the
idealized spectral-type sequence illustrated in Figure 8.
Measured stellar properties---including the dominant NRP velocity
amplitudes---would be used as input constraints for models
similar to those described in this paper.
These models could then produce specific predictions for whether
there should (or should not) be an equatorial disk for each star.
Other models of the Be phenomenon, such as those summarized in
{\S}~1, will likely result in a different division of predicted
disks versus non-disks for the same database of stars.
Comparisons of each set of predictions with the observed
occurrence of disks would act as a clear test of the viability
of the proposed physical processes.

\acknowledgments

I gratefully acknowledge Adriaan van Ballegooijen,
Richard Townsend, and Stan Owocki for many valuable discussions.
I also thank the intrepid and indomitable George Collins
for providing the inspiration for this work.
This research was supported by the National Aeronautics and Space
Administration (NASA) under grant {NNG\-04\-GE77G} to the
Smithsonian Astrophysical Observatory.

\appendix
\section{Derivation of Second Order Wave Pressure Terms}

For simplicity, the equations of hydrodynamic mass and momentum
conservation are written below in Cartesian, plane-parallel
coordinates.
The vertical direction is $z$ and the relevant horizontal direction
for oblique wave propagation is $x$.
This geometry is applied to the equatorial plane of a rotating
system, where $z$ points radially outward and $x$ points in the
prograde azimuthal direction (i.e., increasing $x$ means
increasing $\phi$ in the direction of rotational motion).
We assume symmetry above and below the equatorial plane, so that
all variations in colatitude $\theta$ (or Cartesian coordinate $y$)
are ignored.

The two nontrivial velocity components are defined as
$u = v_{x}$ and $w = v_{z}$.
The equations of mass and momentum conservation are written as
\begin{equation}
  \frac{\partial \rho}{\partial t} +
  \frac{\partial}{\partial x} (\rho u) +
  \frac{\partial}{\partial z} (\rho w) \, = \, 0
\end{equation}
\begin{equation}
  \frac{\partial}{\partial t} (\rho w) +
  \frac{\partial}{\partial x} (\rho u w) +
  \frac{\partial}{\partial z} (\rho w^{2}) +
  \frac{\partial P}{\partial z} - \rho g \, = \, 0
\end{equation}
\begin{equation}
  \frac{\partial}{\partial t} (\rho u) +
  \frac{\partial}{\partial x} (\rho u^{2}) +
  \frac{\partial}{\partial z} (\rho u w) +
  \frac{\partial P}{\partial x} \, = \, 0  \,\, .
\end{equation}
It is important to write the above equations in their full
``conservation form.''
Let us assume that all dynamical variables can be split into the
sum of zeroth order (time steady) and first order (sinusoidally
oscillating) components.
Initially, the radial wind velocity $w_{0}(z)$ and the
mean rotation velocity $u_{0}(z)$ are assumed to have arbitrary
dependences on height.
The mean density scale height is defined as
$H \equiv -\rho_{0} / (\partial \rho_{0} / \partial z)$.
All zeroth order properties are assumed to be constant in both
time $t$ and horizontal position $x$.

Retaining only the zeroth order terms yields equations for the
mean atmospheric stratification in the absence of wave pressure.
Collecting the first order terms yields the dispersion
relation as well as relative amplitude relations for linear waves.
These are discussed by, e.g., Mihalas \& Mihalas (1984) and
Wang et al.\  (1995), and the first-order expressions themselves
are given in equations (7.82)--(7.90) of Cranmer (1996).
The remainder of this Appendix deals with the second order terms;
i.e., those that involve products of two first order fluid
quantities.
If the time average of a given second order term (taken over either
a single wave period of an integer multiple of periods) does not
vanish, then it gives rise to a net contribution to the time-steady
conservation equations.

For example, the period-averaged mass flux conservation equation
can be written as
\begin{equation}
  \frac{\partial}{\partial z} (\rho_{0} w_{0}) +
  \left\langle \frac{\partial}{\partial x} (\rho_{1} u_{1})
  \right\rangle +
  \left\langle \frac{\partial}{\partial z} (\rho_{1} w_{1})
  \right\rangle \, = \, 0  \,\, .
  \label{eq:mass02}
\end{equation}
The period-averaging is denoted by angle brackets.
These terms can be simplified further by exploring some of the
mathematical properties of such time averages.
Consider a complex first order quantity 
$f = (f_{r} + i f_{i}) e^{i (\omega t - k_{x} x)}$, where the
real amplitudes $f_r$ and $f_i$ are assumed to contain an unspecified
dependence on $z$.
For two such quantities ($f$ and $g$), the Eulerian average of their
product over a period $2\pi / \omega$ has the following real part:
\begin{equation}
  \langle fg \rangle \, = \,
  \frac{1}{2} (f_{r}g_{r} + f_{i}g_{i}) \, = \,
  \frac{1}{4} (f^{\ast}g + fg^{\ast})
\end{equation}
where $f^{\ast}$ denotes the complex conjugate of $f$ and this
averaged quantity is constant in $t$ and $x$.
An additional and useful identity can be derived for these kinds
of sinusoidal time variations,
\begin{equation}
  \left\langle \frac{\partial}{\partial t} (fg) \right\rangle
  \, = \,
  \left\langle f \frac{\partial g}{\partial t} \right\rangle +
  \left\langle g \frac{\partial f}{\partial t} \right\rangle
  \, = \, 0  \,\, ,
  \label{eq:fgiden} 
\end{equation}
and a similar identity can be defined by replacing $t$ with $x$,
since the $x$ variations are ideally sinusoidal as well. 
(This cannot be done for the variations in $z$, however.)
Thus, the second term in equation (\ref{eq:mass02}) is zero, and
the radial derivative in the third term can be taken outside the
angle brackets to obtain
\begin{equation}
  \frac{\partial}{\partial z} \left( \rho_{0} w_{0} +
  \langle \rho_{1} w_{1} \rangle \right)  \, = \, 0 \,\, .
  \label{eq:massLag}
\end{equation}   
This is the Lagrangian form of the mass conservation equation,
which indicates that longitudinal and compressive waves can be
responsible for a nonzero mass flux even in the absence of a
mean local (Eulerian) flow velocity $w_0$.
The ramifications of the $\langle \rho_{1} w_{1} \rangle$ term
(also sometimes called the ``Stokes drift'') are discussed
in {\S}~4.

The sum of the second order terms in the $x$ and $z$ components of
the momentum conservation equation form a vector that we define
as ${\bf S} = S_{x} \hat{\bf e}_{x} + S_{z} \hat{\bf e}_{z}$.
Dimensionally, ${\bf S}$ has the units of a pressure gradient,
or force per unit volume, and its components are
\begin{equation}
  S_{x} \, = \, \frac{\partial}{\partial t} (\rho_{1} u_{1}) +
  \frac{\partial}{\partial x}
  (\rho_{0} u_{1}^{2} + 2\rho_{1} u_{0} u_{1}) +
  \frac{\partial}{\partial z}
  (\rho_{0} u_{1} w_{1} + \rho_{1} u_{0} w_{1} + \rho_{1} u_{1} w_{0})
\end{equation}
\begin{equation}
  S_{z} \, = \, \frac{\partial}{\partial t} (\rho_{1} w_{1}) +
  \frac{\partial}{\partial x}
  (\rho_{0} u_{1} w_{1} + \rho_{1} u_{0} w_{1} + \rho_{1} u_{1} w_{0})
  + \frac{\partial}{\partial z}
  (\rho_{0} w_{1}^{2} + 2\rho_{1} w_{0} w_{1})  \,\, .
\end{equation}
These expressions can be simplified by noticing that the derivatives
with respect to $t$ and $x$ can be neglected when constructing the
period averages of $S_x$ and $S_z$.
Thus, the mean azimuthal wave pressure gradient is given by
\begin{equation}
  \langle S_{x} \rangle \, = \,
  \left\langle \frac{\partial}{\partial z} \left(
    \rho_{0} u_{1} w_{1} \right) \right\rangle +
  \left\langle \frac{\partial}{\partial z} \left(
    u_{0} \rho_{1} w_{1} \right) \right\rangle +
  \left\langle \frac{\partial}{\partial z} \left(
    w_{0} \rho_{1} u_{1} \right) \right\rangle  \,\, .
  \label{eq:Sxpent}
\end{equation}
The first and second terms in equation (\ref{eq:Sxpent}) are
referred to by Lee \& Saio (1993) and Lee (2007) as occurring
on transport time scales $\tau_1$ and $\tau_2$, respectively.
The third term depends on the Eulerian mean velocity $w_0$ and
is traditionally ignored in stellar interiors models (for which
the only mean motions are meridional circulations with speeds many
orders of magnitude smaller than the pulsation amplitudes).
We also neglect the $w_0$ term here (see below),
and the remaining terms can be written as
\begin{equation}
  \langle S_{x} \rangle \, = \,
  \frac{\partial}{\partial z} \left(
    \rho_{0} \langle u_{1} w_{1} \rangle \right) +
  \frac{\partial}{\partial z} \left(
    u_{0} \langle \rho_{1} w_{1} \rangle \right)  \,\, .
  \label{eq:Sxfin}
\end{equation}
Because the first term above depends on $u_1$, its overall sign thus
depends on the sign of $k_x$ (see eq.~[\ref{eq:u1w1simp}]).
The second term, though, has the same sign no matter whether the
waves are prograde or retrograde.
For this term, it is straightforward to see that when
$\rho_1$ and $w_1$ are in phase with one another (e.g., for
an upwardly propagating wave) and $u_0$ is positive (i.e., for
bulk rotation in the $+x$ direction), it behaves similarly to the
radial wave pressure.
In other words, if the pressure-like quantity being
differentiated with respect to $z$ decreases as one goes up in
height, it will result in a positive force (i.e., an upward
acceleration for $S_z$ and an acceleration in the prograde
direction for $S_x$).

The vertical (radial) wave pressure gradient can be condensed
down to only two terms, with
\begin{equation}
  \langle S_{z} \rangle \, = \,
  \left\langle \frac{\partial}{\partial z} \left(
    \rho_{0} w_{1}^{2} \right) \right\rangle +
  2 \left\langle \frac{\partial}{\partial z} \left(
    w_{0} \rho_{1} w_{1} \right) \right\rangle  \,\, .
\end{equation}
As above, the term depending explicitly on $w_0$ will be neglected here.
Thus, the single remaining term
\begin{equation}
  \langle S_{z} \rangle \, = \,
  \frac{\partial}{\partial z} \left(
    \rho_{0} \langle w_{1}^{2} \rangle \right)
  \label{eq:Szfin}
\end{equation}
is equivalent to the plane-parallel limit of the wave pressure
gradient derived elsewhere (e.g., Jacques 1977; Koninx 1992) in
the limit of $\gamma = 1$ (isothermal fluctuations).
Also, the above expression is equivalent to
equation (\ref{eq:gwpr0}) as long as the waves are sinusoidal
in shape ($s=2$).

It should be emphasized that the above form of the radial wave
pressure (eq.~[\ref{eq:Szfin}]) is the same as that derived by
Jacques (1977) and others, despite the fact that they do {\em not}
make the assumption that $w_{0} = 0$.
In fact, they apply equation (\ref{eq:Szfin}) directly to the
problem of how waves can help accelerate {\em supersonic winds.}
Thus, it is likely that taking the limit of $w_{0} \rightarrow 0$
in the above equations is equivalent to transforming into the
local comoving frame of reference of the fluid.\footnote{%
In the case of the horizontal (azimuthal) coordinate, this
transformation cannot be made so simply.  Only if it were truly
a plane-parallel system should the mean azimuthal velocity
$u_0$ be set to zero in a similar manner as $w_0$ was set to zero.
For a rotating system, the terms that depend on $u_0$ are likely
to depend on the centrifugal or Coriolis forces, which cannot be
removed by a simple coordinate transformation.}
The proper Reynolds stresses in $S_x$ and $S_z$ are those that add
momentum to the fluid in the comoving (Lagrangian) frame.

\vspace*{0.33in}
\begin{figure}[h]
\epsscale{1.11}
\figurenum{6}
\plotone{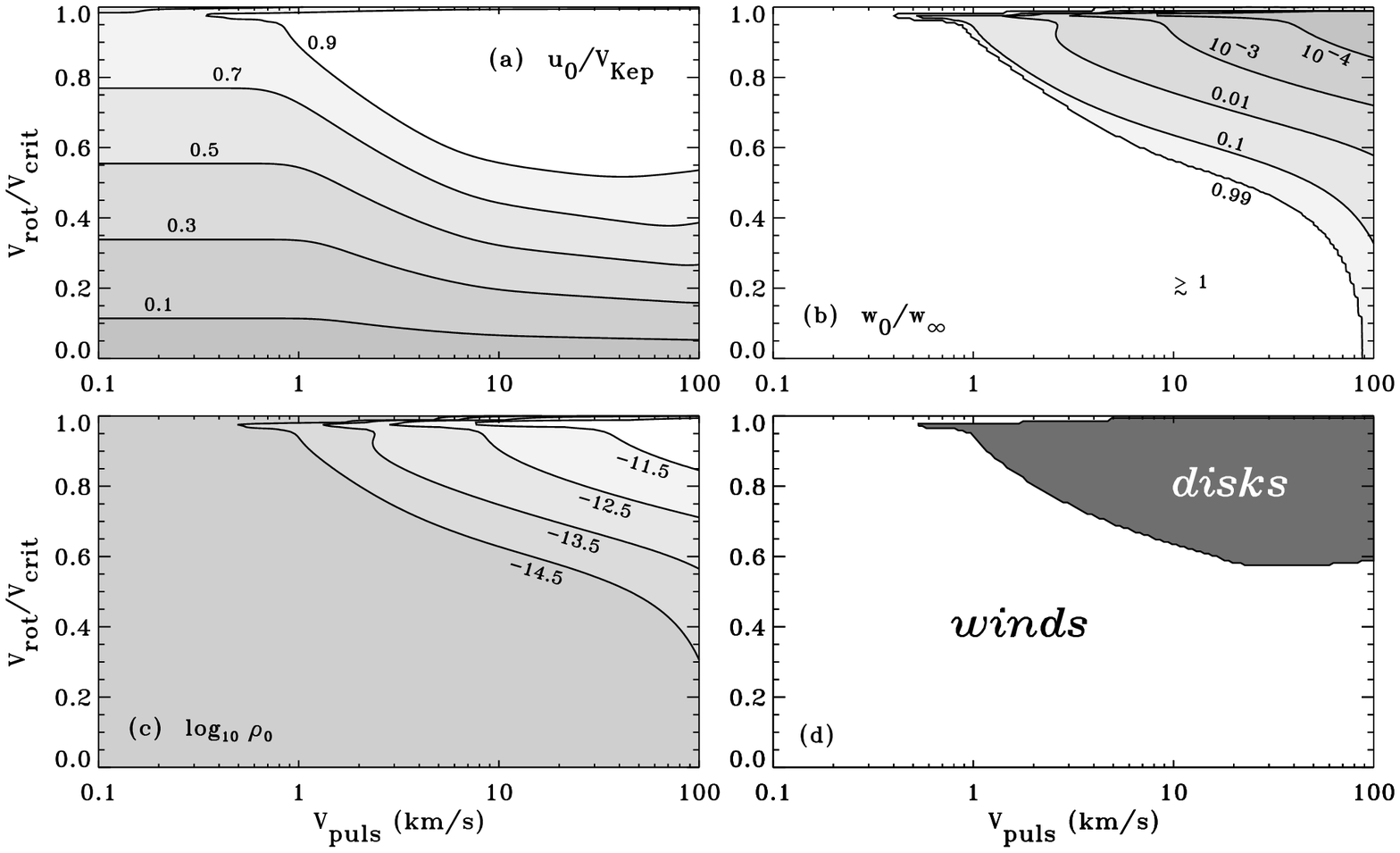}
\caption{Properties at the largest height ($z = 80 H_0$) for
a grid of B2~V star models that vary $V_{\rm rot}$ and
$V_{\rm puls}$ independently of one another:
({\em{a}}) ratio of rotation speed $u_0$ to Keplerian orbital
speed $V_{\rm Kep}$,
({\em{b}}) ratio of radial wind speed $w_0$ to assumed terminal speed
$w_{\infty}$, and
({\em{c}}) mean mass density $\rho_0$.
Labels for constant contour values are given next to each curve.
Panel ({\em{d}}) shows the heuristic dividing line between winds
and disks (see text).}
\end{figure}

\end{document}